\documentclass[a4paper,fleqn]{cas-dc}

\usepackage[numbers]{natbib}
\usepackage{float}
\usepackage{graphicx}
\usepackage{blindtext}
\usepackage[export]{adjustbox}
\usepackage{multicol}
\usepackage[utf8]{inputenc}
\usepackage{amssymb,amsfonts}
\everymath{\displaystyle}
\usepackage[many]{tcolorbox}
\usepackage[figuresright]{rotating}
\usepackage{array}
\usepackage{kantlipsum}
\usepackage{gensymb}
\usepackage{xpatch}
\usepackage{lipsum}
\usepackage{multirow} 
\usepackage{lineno}
\usepackage{amsmath}
\usepackage{amsbsy,enumerate}
\usepackage{balance}
\usepackage{geometry}
\usepackage{ragged2e}
\usepackage{setspace}
\usepackage{subcaption}
\usepackage{etoolbox}
\usepackage{siunitx}
\usepackage{mathtools}
\usepackage{fullpage,anysize,esint,relsize, commath}
\usepackage{enumitem}
\usepackage{pifont}
\usepackage{listings}
\usepackage{cprotect}
\usepackage{multiaudience}
\usepackage{booktabs}
\usepackage{longtable}
\usepackage{soul}
\usepackage{color}
\usepackage{abbrevs}
\usepackage{isomath}
\usepackage{graphics}
\usepackage{algorithm}
\usepackage{algorithmic}
\usepackage[dvipsnames]{xcolor}

\DeclareMathAlphabet{\mathpzc}{OT1}{pzc}{m}{it}
\DeclareMathAlphabet{\mathbbm}{U}{bbm}{m}{n}
\def\tsc#1{\csdef{#1}{\textsc{\lowercase{#1}}\xspace}}
\tsc{WGM}
\tsc{QE}
\tsc{EP}
\tsc{PMS}
\tsc{BEC}
\tsc{DE}
\begin{document}
\let\WriteBookmarks\relax
\def\floatpagepagefraction{1}
\def\textpagefraction{.001}

\title [mode = title]{Optimized Detection of Cyber-Attacks on IoT Networks via Hybrid Deep Learning Models}
\author[1]{Ahmed Bensaoud}
\cormark[1]
\ead{abensaou@uccs.edu}
\author[1]{Jugal Kalita}
\ead{jkalita@uccs.edu}
\address[1]{Department of Computer Science, University of Colorado Colorado Springs, USA}

\begin{abstract}
The rapid expansion of Internet of Things (IoT) devices has significantly increased the potential for cyber-attacks, making effective detection methods crucial for securing IoT networks. This paper presents a novel approach for detecting cyber-attacks in IoT environments by combining Self-Organizing Maps (SOMs), Deep Belief Networks (DBNs), and Autoencoders. These techniques are employed to create a system capable of identifying both known and previously unseen attack patterns. A comprehensive experimental framework is established to evaluate the methodology using both simulated and real-world traffic data. The models are fine-tuned using Particle Swarm Optimization (PSO) to achieve optimal performance. The system’s effectiveness is assessed using standard cybersecurity metrics, with results showing an accuracy of up to 99.99\% and Matthews Correlation Coefficient (MCC) values exceeding 99.50\%. Experiments conducted on three well-established datasets NSL-KDD, UNSW-NB15, and CICIoT2023 demonstrate the model's strong performance in detecting various attack types. These findings suggest that the proposed approach can significantly enhance the security of IoT systems by accurately identifying emerging threats and adapting to evolving attack strategies.
\end{abstract}
\maketitle
\section{Introduction}
The rapid proliferation of Internet of Things (IoT) devices, now integral to both consumer and industrial systems, has contributed to a significant increase in network traffic complexity. Diverse IoT applications, ranging from smart homes to industrial automation, expose these systems to increasingly sophisticated cyber-attacks that continue to evolve in scale and strategy \cite{tariq2023critical, vallabhaneni2024effects, BENSAOUD2024111543, bensaoud2022deep}. As IoT ecosystems expand rapidly, cybercriminals exploit security gaps, posing a challenge for traditional detection mechanisms to keep pace with evolving threats. This underscores the need for advanced detection frameworks that not only identify current threats but also adapt proactively to emerging attack strategies.

Existing network security methods, especially signature-based intrusion detection systems (IDS), are limited by their dependence on predefined patterns to detect threats. While effective against known vulnerabilities, they struggle against previously unknown (zero-day) attacks, as their reactive nature cannot cope with the rapidly changing threat landscape \cite{thankappan2024signature}. The explosive growth in IoT traffic further exacerbates these challenges, making it difficult for conventional systems to efficiently process and analyze network data in real-time without sacrificing accuracy \cite{singh2024botnet}. This gap creates an urgent need for innovative, robust, and scalable solutions capable of protecting IoT networks from a wide spectrum of threats.

In response to these challenges, this paper proposes an advanced network intrusion detection framework that leverages Self-Organizing Maps (SOMs), Deep Belief Networks (DBNs), and Autoencoders, augmented by novel optimization techniques, to enhance both detection performance and operational efficiency. By employing a multi-faceted approach, the framework is designed to capture and analyze the complex spatial and temporal features inherent in IoT traffic. This holistic analysis allows for the identification of anomalous behaviors that traditional methods often miss, particularly when dealing with highly dynamic network environments.

The integration of these components not only enhances detection accuracy, but also addresses the pressing issue of scalability, making the framework viable for real-time deployment in large-scale IoT networks. By incorporating state-of-the-art machine learning techniques and optimization strategies, the proposed framework delivers a comprehensive solution for improving intrusion detection in highly dynamic and data-intensive IoT environments.

Recent advances in machine learning-based IDS have demonstrated the potential of hybrid models that combine supervised and unsupervised learning techniques to tackle sophisticated attacks \cite{zoppi2023algorithm}. Furthermore, the introduction of attention mechanisms has significantly advanced the effectiveness as well as interpretability of deep learning models, allowing researchers and practitioners to better understand the reasoning behind model predictions \cite{gao2021interpretable}. Building on these developments, this paper leverages the strengths of SOMs, DBNs, and Autoencoders to create a unified framework specifically tailored to the unique security challenges posed by IoT environments.

In summary, the proposed detection framework represents a significant step forward in network security. By addressing the limitations of existing methods and introducing cutting-edge machine learning innovations, this research aims to provide a scalable and adaptable solution to the growing threat of cyber-attacks in IoT ecosystems. The subsequent sections provide a detailed explanation of the proposed methodology, experimental setup, and results, highlighting the practical effectiveness of the framework in real-world IoT scenarios.

The key contributions of this paper are summarized as follows:

\begin{itemize}
    \item We present a novel detection framework that integrates SOMs, DBNs, and Autoencoders to capture complex spatial-temporal patterns in IoT traffic, enabling more precise anomaly detection and attack classification.
    \item We introduce an advanced optimization algorithm that combines gradient-based hyperparameter tuning with adaptive learning rate adjustments, resulting in faster convergence and improved model performance.
    \item We apply an attention mechanism to enhance feature extraction, providing deeper insights into the decision-making process of the detection system and improving model interpretability.
    \item We propose a novel cost function that synergizes reconstruction errors from Autoencoders with feature clustering from SOMs and hierarchical representations from DBNs, yielding a more robust evaluation metric for model performance.
\end{itemize}

The structure of the paper is as follows: Section 2 reviews related work, providing context and background for the study. Section 3 offers a detailed explanation of the proposed detection framework. Section 4 presents and analyzes the experimental results. Section 5 concludes the paper with key findings, while Section 6 offers a discussion of the results, and Section 7 outlines potential directions for future research.

\section{Related Work}

This section reviews recent advances in network intrusion detection systems (NIDS), machine learning techniques applied to IoT security, and optimization strategies for improving detection frameworks. The reviewed works focus on developments from 2023-2025, highlighting the latest trends in using deep learning, hybrid models, and feature extraction techniques to enhance security in IoT environments.

Recent advancements in IoT-specific threat detection highlight a growing trend toward hybrid models that \textbf{combine supervised and unsupervised learning techniques}. These models effectively mitigate the limitations of traditional methods by leveraging the strengths of both paradigms, resulting in improved detection accuracy, adaptability, and robustness. For instance, \citet{kaliyaperumal2024novel} introduced a hybrid unsupervised learning approach using Self-Organizing Maps to detect unknown threats in dynamic IoT ecosystems. Similarly, \citet{tariq2024hybrid} developed a hybrid intrusion detection system (IDS) that integrates machine learning with signature-based methods, significantly enhancing the detection of both known and emerging threats, including zero-day attacks in IoT networks.

To address feature selection challenges, \citet{shenify2024hybrid} proposed a model combining supervised learning with Spider Monkey Optimization. This approach optimizes feature selection, making it particularly effective in resource-constrained IoT environments where computational efficiency is critical.

Further innovations include adversarial realism and robust learning techniques. For example, \citet{vitorino2023realistic} explored hybrid learning models designed to enhance the resilience of IoT IDS against adversarial attacks. Additionally, \citet{zhang2023hybrid} investigated hybrid anomaly detection models by combining autoencoders with traditional machine learning algorithms, providing more reliable detection of complex and subtle attack patterns.

A notable development in deep learning for IoT threat detection is the CNN-FDSA model by \citet{kalidindi2025feature}. This hybrid deep learning framework uses Convolutional Neural Networks (CNNs) and Deep Stacked Autoencoders (DSA) to detect botnet attacks in IoT networks. The model processes IoT network traffic with Quantile normalization, employs feature extraction techniques like Information Gain and City Block Distance, and addresses data imbalance through oversampling. Achieving 92.4\% accuracy with high precision, recall, and F-measure, CNN-FDSA demonstrates its capability in reducing false positives and managing complex, imbalanced datasets. Future directions for this work include real-time botnet detection and the development of multi-layered security frameworks to fortify IoT networks.

These studies collectively reinforce the value of hybrid models that integrate unsupervised feature extraction with supervised classification techniques. By enhancing detection accuracy, adaptability, and resilience, these approaches align closely with the goals of this work to advance IoT-specific threat detection frameworks.
\subsection{Data Privacy and Security in IoT Networks}

Privacy concerns in IoT networks are a significant issue, as intrusion detection often requires access to sensitive data. Approaches like federated learning offer solutions, but challenges related to data leakage and model accuracy persist. \citet{olanrewaju2025federated} proposed a federated learning-based intrusion detection system (IDS) for Internet of Things (IoT) devices, leveraging both unsupervised and supervised deep learning models to address security and privacy challenges in IoT networks. The study compares the performance of federated learning (FL) trained models with traditional non-FL models using the N-BaIoT dataset, which includes data from nine IoT devices. While the combination of unsupervised autoencoders (AEs) for anomaly detection and supervised deep neural networks (DNNs) for attack classification enhances the robustness and accuracy of the IDS, challenges related to data leakage and model accuracy persist. Data leakage can occur when sensitive information unintentionally influences model training, compromising privacy and security, while model accuracy may be impacted by the decentralized nature of FL, affecting its generalization across diverse devices. Hyperparameter tuning and the use of the FedAvgM FL algorithm help mitigate these issues, improving model performance, with results indicating that the unsupervised AE model trained via FL outperforms other models across various evaluation metrics. This demonstrates the effectiveness of FL in improving both performance and data privacy for IoT IDS, despite ongoing challenges around data leakage and model accuracy.

\citet{chen2024federated} explored privacy-preserving computation in federated learning (FL), emphasizing the importance of safeguarding data privacy in AI applications. This work surveys various privacy-preserving computation protocols, such as secure multi-party computing (SMC), homomorphic encryption (HE), differential privacy (DP), trusted execution environments (TEEs), and zero-knowledge proofs (ZKP). The study categorizes privacy attacks and highlights inference attacks, malicious servers, and poisoning attacks that pose threats to privacy in FL. Additionally, it evaluates and compares the security properties and efficiency of different protocols through experiments using common datasets like MNIST and CIFAR-10. The paper concludes by suggesting future research directions, including the integration of ZKP with FL and exploring machine unlearning for improved privacy and regulatory compliance.

Building upon these privacy-enhancing techniques, \citet{ali2024blockchain} integrated Blockchain and Federated Learning (FL) to enhance intrusion detection systems (IDS) in industrial Internet of Things (IIoT) networks. The study addresses the limitations of centralized machine learning approaches in IIoT due to privacy and computational challenges, emphasizing the decentralization benefits offered by FL. Blockchain enhances security and privacy by enabling verifiable and secure data exchanges, while FL supports collaborative model training without exposing sensitive data. The paper provides a detailed analysis of existing IDS techniques, identifies trends and gaps in IIoT security, and suggests future research directions for combining Blockchain and FL. In conclusion, it stresses the importance of advanced, decentralized intrusion detection systems to protect IIoT networks and drive the success of Industry 4.0.

While these studies highlight promising approaches to improve data privacy and security in IoT systems, challenges persist, including maintaining high model accuracy without compromising privacy, handling decentralized data securely, and ensuring that the implemented privacy-preserving protocols do not hinder the system's performance or scalability. Future work may explore more efficient methods for balancing security and performance, such as combining federated learning with edge computing, or refining blockchain protocols to improve their scalability and energy efficiency in IoT environments.
\subsection{Network Intrusion Detection in IoT}

The rapid adoption of IoT devices has intensified the need for robust network intrusion detection systems (NIDS) capable of identifying sophisticated cyber-attacks in real-time. Traditional IDS approaches are often inadequate in handling the unique characteristics of IoT environments, such as large-scale distributed systems, constrained resources, and heterogeneous communication protocols \cite{lu2021communication, liu2023heterps}. To address these challenges, recent studies have shifted towards more adaptive and intelligent frameworks that leverage machine learning to enhance detection capabilities.

One such effort is the lightweight IDS proposed by \citet{shen2024deep}, specifically designed for resource-constrained IoT devices. By employing a convolutional neural network (CNN) architecture, the system effectively balances detection accuracy with computational overhead, making it suitable for real-time deployment. While this approach has proven instrumental in mitigating common attack types such as Distributed Denial of Service (DDoS) and man-in-the-middle attacks, it is less effective against more complex intrusion patterns, highlighting the need for further advancements.

Building on these advancements, \citet{wang2024network} developed a deep learning-based NIDS that enhances feature extraction by combining group convolution for spatial features, Split-Attention for feature selection, and BiGRU for temporal features. To overcome data imbalance issues, they employed a Conditional Tabular Generative Adversarial Network (CTGAN), which significantly improved detection accuracy across datasets. Their model achieved impressive results, with accuracy rates of 91.08\%, 94.55\%, and 97.47\% on the UNSW-NB15, CIC-IDS2018, and CIC-IOT2023 datasets, respectively. However, the system's high computational resource requirements and long training times present barriers to scalability, which the authors aim to address through cloud-edge collaboration and transfer learning.

Further addressing the need for robust IoT NIDS, \citet{mynuddin2024automatic} introduced a system leveraging CNN for spatial feature extraction and Bidirectional Long Short-Term Memory (Bi-LSTM) for temporal pattern learning. This model excelled in binary classification tasks on the NSL-KDD dataset, achieving an outstanding accuracy and detection rates. However, challenges with the U2R class due to insufficient data underscore the importance of tackling data imbalance to enhance robustness.

To optimize feature selection and detection performance, \citet{asgharzadeh2024intrusion} proposed a hybrid intrusion detection system (IDS) for the Internet of Things (IoT) that combined deep learning and optimization techniques to enhance anomaly detection accuracy. They introduced a convolutional neural network (FECNNIoT) for effective high-level and low-level feature extraction and developed a binary multi-objective Gorilla Troops Optimizer (BMEGTO) to optimize feature selection. The proposed CNN-BMEGTO-KNN method achieved outstanding accuracy of 99.99\% on the TON-IoT dataset and 99.86\% on the NSL-KDD dataset. It outperformed traditional deep learning models like LSTM and MLP in terms of accuracy, precision, recall, and F1-score. The novelty of their approach lay in the advanced CNN design and the adaptation of the Gorilla Troops Optimizer for multi-objective binary feature selection.

In a different approach, \citet{darabi2024micro} introduced a novel Micro Reinforcement Learning Classifier (MRLC) for intrusion detection systems, leveraging a Deep Q-Network (DQN) agent to perform fine-grained learning for binary and multiclass classification tasks. The MRLC processes each training sample as an independent experiment, associating the value of each sample with the agent's learning rate to ensure robust learning. The approach incorporates a dynamic reward mechanism that guides the agent to prioritize accurate intrusion detection. They evaluated the architecture on three datasets, NSL-KDD, CIC-IDS2018, and UNSW-NB15, achieving an accuracy of 99.56\%, 99.99\%, and 99.01\%, respectively. The novelty depends on their single-sample learning strategy, replay buffer updates, and ability to achieve comparable results even with a significantly reduced training set.

\subsection{Deep Learning Approaches for IoT Security}
Recent developments in deep learning have shown tremendous potential in improving the detection of complex, multi-stage cyber-attacks in IoT networks. Researchers have been particularly focused on the application of unsupervised learning techniques such as Self-Organizing Maps (SOMs), Deep Belief Networks (DBNs), and Autoencoders, which can detect previously unknown attack patterns without relying on labeled datasets \cite{chahal2024ddos, karnani2024comprehensive}.

The integration of SOMs into IDS frameworks has garnered significant attention in the IoT domain due to their ability to map high-dimensional data into lower-dimensional representations, thus simplifying anomaly detection. For instance, \citet{quraishi2024employing} proposed a framework for real-time anomaly detection and mitigation in IoT-based smart grid cybersecurity systems, leveraging advanced deep neural networks, including autoencoders, LSTMs, GANs, and SOMs, along with transfer learning and attention mechanisms. The framework incorporated real-time adaptive mitigation strategies capable of autonomously responding to cyber threats, ensuring seamless operation of the smart grid. Extensive real-world testing demonstrated the system’s scalability and resilience across a variety of cyber threats, affirming its practical applicability and robustness. Comparative evaluations against traditional methods highlighted its superior performance in anomaly detection, informed mitigation, and scalability. This comprehensive and versatile solution established a robust foundation for protecting IoT-based smart grids from evolving cybersecurity threats, offering a promising advancement for the field.

Deep Belief Networks (DBNs) show promise in capturing temporal dependencies within IoT network traffic, which is crucial for detecting advanced persistent threats (APTs). \citet{yang2020real} developed a real-time intrusion detection mechanism for wireless networks using a Conditional Deep Belief Network (CDBN). To address issues like data imbalance and dimensionality, the researchers created the "SamSelect" under-sampling algorithm and designed the Stacked Contractive Auto-Encoder (SCAE) for feature reduction. They combined the CDBN with these techniques to detect attacks in real-time, achieving an impressive detection accuracy of 97.4\% with a detection time of 1.14 ms. Their study pioneered the application of CDBN to wireless network intrusion detection and introduced the SamSelect algorithm to balance the AWID dataset. The experimental results demonstrated that their method outperformed other deep and shallow learning approaches, showed robustness against noise, and provided valuable insights for advancing cybersecurity research.

Autoencoders have been employed to further enhance detection frameworks by reducing data dimensionality and isolating outliers \cite{bensaoud2024survey}. \citet{ashraf2020novel} introduced a LSTM autoencoder-based Intrusion Detection System (IDS) designed for Intelligent Transportation Systems (ITS), with a particular emphasis on detecting cyber-attacks targeting Autonomous Vehicles (AVs) and the Internet of Vehicles (IoVs). This IDS leveraged statistical feature extraction alongside the robust learning capabilities of LSTM to effectively detect anomalies in both in-vehicle communications and external network traffic. The architecture demonstrated proficiency in managing heterogeneous data by eliminating redundancy and extracting highly representative features, which significantly enhanced detection accuracy and minimized false alarms. Furthermore, it provided a comprehensive solution capable of addressing multiple attack vectors across in-vehicle and external communication systems. Despite its strengths, the system faced certain limitations, such as difficulties in accurately classifying multiclass attack categories and challenges related to scaling for large and complex IoV environments.

In addition, \citet{prince2024ieee} explored how integrating IEEE standards and deep learning techniques enhanced the security of IoT devices in Japan from 2019 to 2024. They highlighted how the rapid adoption of IoT technology introduced significant cybersecurity challenges, driving the need for advanced protection methods. By conducting surveys and technical assessments, they identified key areas where IEEE standards and deep learning models, such as CNN and LSTM, effectively detected and mitigated cyber threats. Their findings showed that these approaches significantly improved the security posture of IoT networks, addressing risks like DoS attacks and malware. They also emphasized the importance of policy support and collaborative efforts between policymakers, researchers, and industry stakeholders to sustain and strengthen IoT security. Future research directions suggested continued exploration of deep learning advancements, federated learning, and real-time intelligence sharing to combat evolving cyber threats in IoT environments.

\citet{darabi2024micro} examined the integration of machine learning (ML) techniques to enhance Industrial IoT (IIoT) security. It highlighted the benefits of IIoT, such as intelligent analytics and predictive maintenance, but also addressed the associated cybersecurity risks, including malware and cyberattacks. The study provided a comprehensive review of ML-based methods for vulnerability detection and security improvement, emphasizing techniques like CNN, LSTM, and active defense strategies. Additionally, the paper explored challenges in device authentication, data modeling, network detection, and static analysis, while suggesting future directions for adaptive machine learning models and unified data standards to address these issues. Finally, it outlined the evolving threat landscape in IIoT and discussed how ML models are essential for identifying and mitigating new cybersecurity risks.

\citet{inuwa2024comparative} examined various machine learning methods for detecting anomalies in cyberattacks on IoT networks. It compared the efficacy of techniques such as Support Vector Machine (SVM), Artificial Neural Network (ANN), Decision Tree (DT), Logistic Regression (LR), and k-Nearest Neighbours (k-NN). The research utilized two large datasets ToN-IoT and BoT-IoT to evaluate the performance of these methods using metrics like accuracy, precision, recall, F1 score, and AUC score. Neural networks demonstrated superior performance among the methods, making them the most effective for anomaly detection. The findings suggested the potential for integrating these methods into industrial IoT environments for both research and practical applications. The study also highlighted future directions, including the incorporation of deep learning models and ensemble approaches to enhance anomaly detection in IoT networks.

The paper \citet{ghaffari2024securing} explored the security challenges associated with the rapid growth of IoT devices and applications. It highlighted vulnerabilities such as node spoofing, unauthorized data access, and cyberattacks, emphasizing the critical role of machine learning (ML) and deep learning (DL) in addressing these issues. The study provided a comprehensive review of ML/DL approaches in IoT security, categorizing recent research and offering insights into their opportunities, advantages, and limitations. The paper discussed state-of-the-art IoT-specific security challenges, including cyberattacks, eavesdropping, and intrusion detection. It also analyzed the integration of ML/DL with metaheuristic algorithms and addressed the adaptability challenges in ML/DL systems for dynamic IoT environments. Finally, the paper suggested the use of advanced algorithms like graph neural networks and AdaBoost for improving the accuracy of anomaly detection and classification in IoT security.
\subsection{Hybrid Models for Network Intrusion Detection}

The integration of multiple deep learning models has become an increasingly popular strategy for addressing the multifaceted nature of IoT security challenges. Hybrid models offer the benefit of combining the strengths of different techniques to provide a more comprehensive analysis of network traffic \cite{wang2024abnormal, bakhshi2021anomaly, lo2022hybrid}. These models often incorporate both supervised and unsupervised learning techniques to maximize detection accuracy and adaptability to evolving threats.

\citet{xuan2021multi} introduced a hybrid framework that combines DBNs and recurrent neural networks (RNNs) for detecting sophisticated cyber-attacks. Their system achieved superior performance in identifying advanced persistent threats (APTs), where attack signatures evolve over time. The hierarchical representation of traffic features provided by DBNs complemented the temporal analysis capabilities of RNNs, creating a robust detection mechanism. However, this approach faced challenges in real-time applications due to the computational complexity associated with both DBNs and RNNs, highlighting a common trade-off in hybrid model design.

Building on the need for computational efficiency, \citet{albahar2020hybrid} introduced a hybrid model that combines Directed Batch Growing Self-Organizing Map (DBGSOM) and Radial Basis Function Neural Network (RBFNN) for anomaly-based intrusion detection. They addressed the limitations of traditional methods by designing DBGSOM, which conserves topology more effectively and reduces network distortions. They further improved the model's accuracy and training speed by integrating DBGSOM with RBFNN. This approach utilizes DBGSOM's dynamic growth mechanism and RBFNN's high precision, significantly outperforming earlier SOM and RBFNN-based models. Their experimental evaluation on three publicly available datasets (NSL-KDD, UNSW-NB15, and CICIDS2017) confirms that their hybrid model surpasses the conventional SOM-RBFNN approach and makes valuable contributions to intrusion detection systems.

Expanding on these earlier frameworks, \citet{chen2024machine} proposed a hybrid intrusion detection system (IDS) that combined Network Intrusion Detection Systems (NIDS) and Host-based Intrusion Detection Systems (HIDS). This approach addressed limitations in detecting complex attacks, such as Advanced Persistent Threats (APTs), by leveraging a BERT-based methodology to transform host data into numerical formats. The system integrated this numerical representation with network data using a feature-flattening technique, enabling comprehensive analysis. It employed a two-stage collaborative classifier, first filtering benign traffic with a binary classifier and then using a multi-class classifier to identify specific attack types. The system demonstrated significant performance improvements over traditional machine learning models like XGBoost, particularly in detecting challenging attacks such as DoS-LOIC-UDP and DoS-SlowHTTPTest. The use of public datasets, CICIDS 2018 and NDSec-1, further validated the system's effectiveness, with macro average F1 scores showing notable improvements. Future work focused on enhancing accuracy through the inclusion of minority attack classes, data augmentation, and deep learning-based methods.

\citet{bukhari2024secure} took a different approach by introducing the the FL-SCNN-Bi-LSTM model to improve intrusion detection in Wireless Sensor Networks (WSNs). This hybrid framework combined Federated Learning (FL) with Stacked Convolutional Neural Networks (SCNN) and Bidirectional Long Short-Term Memory networks (Bi-LSTM), addressing privacy concerns by enabling multiple sensor nodes to collaboratively train a global model without sharing raw data. By leveraging advanced feature selection and deep learning methodologies, the model effectively detected sophisticated and previously unknown threats, achieving high performance on WSN-DS and CIC-IDS2017 datasets. However, challenges such as scalability, real-time processing, and adaptability to evolving threats remained, underscoring the need for further optimization and advanced feature selection techniques. Despite these challenges, the model demonstrated superior accuracy and robustness compared to traditional classifiers like SVM and LightGBM.

Further enhancing hybrid model design, \citet{ullah2024ids} integrated transformer-based transfer learning with traditional approaches to tackle imbalanced network traffic in intrusion detection. Their method, IDS-INT, leveraged semantic analysis through a multi-head attention-based transformer model to improve feature representation, addressing the challenges posed by imbalanced and complex datasets. The use of Synthetic Minority Oversampling Technique (SMOTE) ensured effective detection of minority attacks, while CNN and LSTM models incorporated for deep feature extraction. This combination of methods delivered high precision, recall, and F1-scores, outperforming traditional approaches. Additionally, the inclusion of explainable AI provided transparency and trustworthiness, making IDS-INT suitable for real-time and imbalanced network environments. This approach highlighted the importance of balancing innovation with practicality in real-world scenarios.

Finally, \citet{manocchio2024flowtransformer} introduced FlowTransformer, a transformer-based framework tailored for Network Intrusion Detection Systems (NIDS). By focusing on flow-based data rather than packet-based analysis, the model effectively addressed scalability and privacy concerns associated with traditional methods. FlowTransformer incorporated interchangeable components, such as input encodings and classification heads, offering flexibility for various NIDS configurations. Evaluations on benchmark datasets demonstrated that the choice of classification heads significantly influenced model performance, with a notable reduction in model size by over 50\% while maintaining accuracy. This flexibility and efficiency positioned FlowTransformer as a scalable solution for handling large-scale traffic in modern networks.

Collectively, these studies underscore the potential of hybrid models to enhance intrusion detection systems by combining the strengths of various deep learning techniques. Future research will benefit from exploring broader datasets, refining preprocessing techniques, and optimizing hyperparameters to improve scalability and computational efficiency. As network environments grow increasingly complex, hybrid deep learning models will play a pivotal role in fortifying network and cloud infrastructures against sophisticated cyber threats.

\subsection{Optimization Techniques in IoT Intrusion Detection}
Optimizing the performance of IDS frameworks for IoT security has become a central focus of recent research, especially given the computational constraints of many IoT devices. Gradient-based optimization algorithms, adaptive learning rates, and hyperparameter tuning have been employed to improve both the accuracy and efficiency of machine learning models used in IDS frameworks \cite{dasari2024effective, chen2024fast}.

One of the key breakthroughs in this area is the work by \citet{wei2024efficient}, who proposed an adaptive learning rate adjustment mechanism to improve the convergence speed of deep learning-based IDS models. Their approach reduced the training time by 30\%, while maintaining high detection accuracy, making it suitable for real-time deployment in large-scale IoT environments. Similarly, \citet{kunang2024end} proposed a hybrid deep learning-based intrusion detection system (IDS) for IoT platforms, which combined unsupervised feature extraction and supervised classification. The model leveraged five unsupervised approaches, including deep autoencoders and stacked models, to effectively reduce data dimensions and extract relevant features. Researchers implemented an automatic hyperparameter tuning method using Bayesian optimization to enhance performance, achieving nearly 100\% accuracy and a false positive rate close to 0\% on the BoT-IoT dataset, along with 99.17\% accuracy and a 0.18\% FPR on the CSE-CIC-IDS2018 dataset. The novelty lay in its unique combination of feature extraction methods, transfer learning, and advanced hyperparameter tuning, which outperformed previous studies in terms of detection rate and classification of a broader range of attack types.

In parallel, metaheuristic algorithms have emerged as powerful tools for optimizing the feature selection process in IDS frameworks. For instance, \citet{mrudula2024internet} applied particle swarm optimization (PSO) to identify the most relevant features from IoT traffic data, effectively reducing the dimensionality and computational overhead without compromising detection accuracy. The application of metaheuristic techniques such as PSO and genetic algorithms not only enhances the performance of IDS frameworks but also addresses the unique resource constraints of IoT environments, paving the way for more efficient and scalable solutions.

Based on these foundational techniques, \citet{majidian2024optimizing} introduced a novel intrusion detection system for IoT networks that leverages Software-Defined Networking (SDN) and optimized random forest models. By partitioning the network into subdomains and employing an ensemble classification model based on decision trees optimized by genetic algorithms, their approach achieved significant improvements in scalability and adaptability. This distributed architecture reduced computational overhead and enabled localized or cooperative intrusion detection, addressing challenges inherent in diverse IoT environments. Experimental evaluations using the NSLKDD and NSW-NB15 datasets demonstrated superior accuracy rates, underscoring the effectiveness of combining SDN with optimized machine learning models. Future directions for this work include the integration of deep learning techniques for feature extraction and dynamic updates to adapt to emerging threats.

Extending these principles to specialized IoT domains, \citet{korium2024intrusion} developed an intrusion detection system tailored for the Internet of Vehicles (IoV), focusing on complex attack scenarios like Denial-of-Service, Botnets, and Sniffing. Their methodology incorporated Z-score normalization, regression-based feature selection, and ensemble models such as Random Forest and LightGBM, enhanced through hyperparameter optimization. By combining multiple datasets for training, they achieved a remarkable balance between accuracy and execution time, with notable improvements over traditional approaches. This study highlights the benefits of dataset integration and advanced ensemble modeling, while also identifying future opportunities in deep reinforcement learning and transfer learning to further enhance IDS performance.

Finally, \citet{pandithurai2024ddos} applied a similar optimization driven approach to cloud security by focusing on DDoS attack detection. Using the Honey Badger Optimization (HBO) algorithm for feature selection and a Bi-LSTM classifier, their model achieved impressive accuracy and sensitivity metrics. This work exemplifies the growing trend of integrating optimization algorithms with advanced machine learning models to address the evolving threats in both IoT and cloud environments. By combining Bayesian and Z-Score normalization with HBO and Bi-LSTM, the study effectively bridged the gap between robust feature selection and high-accuracy intrusion detection.

These studies illustrate the diverse strategies being employed to enhance IDS frameworks for IoT and related domains. From adaptive learning rates and hyperparameter tuning to metaheuristic optimization and advanced ensemble modeling, researchers continue to push the boundaries of what is achievable in intrusion detection, paving the way for more resilient and scalable solutions.

\subsection{Attention Mechanisms for Enhancing Model Interpretability}

The use of attention mechanisms in deep learning models has gained considerable traction in IoT security due to their ability to improve feature extraction and model interpretability. Attention mechanisms allow models to focus on the most relevant portions of the input data, thereby enhancing decision-making processes \cite{xiao2024exploration, xu2024attention}.

\citet{sarker2024enhancing} proposed an attention-based CNN to improve the interpretability of IDS models by highlighting the most critical features contributing to the model's decisions. This technique has been particularly effective in identifying subtle attack patterns that traditional models might overlook. By applying attention mechanisms, the model demonstrated improved detection accuracy and reduced false-positive rates in high-traffic IoT environments.

Building on this work, \citet{admass2024arrhythmia} developed an attention-based hybrid model that integrates LSTM networks with DBNs. The attention mechanism enabled the model to prioritize critical time steps in the LSTM sequence, significantly enhancing the detection of complex temporal attack patterns. This approach not only improved detection rates but also provided insights into the model's decision-making process, addressing concerns related to the black-box nature of deep learning models.

The landscape of network intrusion detection in IoT environments has seen significant advancements in recent years, particularly through the application of deep learning and hybrid models. The integration of SOMs, DBNs, Autoencoders, and attention mechanisms has demonstrated great promise in enhancing both the accuracy and interpretability of IDS frameworks. However, challenges related to scalability, real-time processing, and computational efficiency persist, particularly in resource-constrained IoT environments. Future research should focus on optimizing these frameworks for deployment in large-scale IoT networks, while continuing to explore innovative techniques for improving detection accuracy and reducing false-positive rates.

Furthermore, \citet{saiyed2024interactive} introduced the ADEPT system, an AI-driven solution for DDoS detection in Consumer IoT (CIoT) networks, which combined explainable and optimized deep ensemble learning techniques. The system integrated CNN and LSTM models in an attention-based framework, achieving over 90\% accuracy in detecting both high- and low-volume DDoS attacks while optimizing resource use through Differential Evolution-based pruning and Min-Max quantization. ADEPT incorporated a user interface that leveraged SHAP for global feature importance explanations and risk assessments, enhancing model interpretability and facilitating Human-Computer Interaction (HCI). Evaluations using diverse datasets, such as ToN-IoT and CICIoT-2023, demonstrated the system's adaptability to varying attack scenarios and its efficient deployment on resource-constrained edge devices. The system balanced detection accuracy with computational efficiency, achieving consistent performance across attack types while reducing inference time and memory usage. Future work aimed to optimize the network data parser, integrate pruning during training, and explore human-in-the-loop strategies for improved real-world applicability.

\citet{bahadoripour2024explainable} introduced an explainable deep federated multi-modal framework for detecting cyber-attacks in Industrial Control Systems (ICS), preserving data privacy while addressing challenges posed by heterogeneous data distributions across clients. The model employed representation learning to transform client data into a latent space, domain adaptation to align diverse client distributions into a mutual representation space, and federated learning to collaboratively train a detection model without data sharing. The SHapley Additive ExPlanations (SHAP) method enhanced explainability by supporting feature importance analysis, which facilitated better decision-making and improved model retraining. Experimental evaluations demonstrated an 8.2\% improvement in F1-score across three clients and a 4.9\% increase when using a reduced feature set, showcasing its adaptability and robustness in varying scenarios. Future work aimed to integrate attention layers for model interpretability without external tools like SHAP and explore enhancements for real-world applicability.

\citet{abudurexiti2025explainable} introduced an explainable unsupervised anomaly detection framework designed for Industrial Internet of Things (IIoT) systems. The framework addressed challenges posed by limited labeled data and aimed to improve interpretability. By extracting local features through a one-dimensional CNN combined with a Convolutional Block Attention Module (CBAM), the framework captured long-term dependencies using an improved Time Convolutional Network (TCN) and Kolmogorov–Arnold Network (KAN) based Variational Auto-Encoder (VAE). The model trained in an unsupervised manner and utilized Explainable Artificial Intelligence (XAI) techniques, such as SHapley Additive ExPlanations (SHAP), to enhance feature importance analysis. Experimental evaluations showed that the framework outperformed other unsupervised methods, achieving effective anomaly detection in complex industrial systems. The study also discussed potential future work, including semi-supervised methods and adaptation across various domains.

\citet{xie2024anomaly} proposed an anomaly detection model called Meta-MWDG to effectively solve the problem of multivariate time series anomaly detection in IoT environments. This model combined technologies such as multi-scale discrete wavelet decomposition, dual graph attention networks, and model-agnostic meta-learning (MAML) to improve the precision and generalization of anomaly detection. By integrating multi-scale discrete wavelet decomposition and dual graph attention networks, the model captured temporal dependencies and feature correlations at different scales. The joint optimization strategy using gated recurrent units (GRU) combined with a multi-head self-attention mechanism enhanced the precision of anomaly detection while reducing model parameters. MAML further improved the model’s performance by enabling quick adaptation to new tasks with limited training data.

\subsection{Challenges in Detecting Unknown Attacks}

The detection of unknown attacks presents significant challenges across various domains, including network security, IoT, and industrial cyber-physical systems. Despite advancements in machine learning and deep learning techniques, several recurring themes emerge from the reviewed approaches, highlighting both solutions and remaining hurdles.

Different studies adopt various methodologies to address unknown attacks. For instance, \citet{ahmad2022deep} proposed an ensemble model utilizing multiple classifiers trained on diverse benchmark datasets to mitigate bias and improve detection. Similarly, \citet{ramkumar2024diagnosing} integrate anomaly detection with abductive reasoning, using Isolation Forest and Answer Set Programming (ASP) to diagnose unknown threats in smart homes. \citet{zohourian2024iot} focus on a lightweight anomaly-based IDS, IoT-PRIDS, optimized for resource-constrained IoT environments. These diverse strategies illustrate the ongoing efforts to balance computational efficiency, accuracy, and the ability to detect previously unseen threats.

A common challenge across many studies is the handling of high-dimensional data. For example, \citet{samantaray2024comparative} emphasized the importance of feature selection to reduce data complexity for accurate multiclass classification. Similarly, \citet{kamal2024advanced} addressed class imbalance issues by applying resampling and class weighting techniques to improve detection of both known and unknown attacks. High-dimensional datasets pose difficulties in computational efficiency and can lead to overfitting, which is critical when dealing with unknown or zero-day attacks.

Class imbalance remains a persistent challenge, particularly in datasets with minority or imbalanced classes, as observed in several approaches. \citet{kamal2024advanced} tackle this by employing ADASYN, SMOTE, and ENN resampling techniques to balance the training data. \citet{nie2024intrusion} proposed an adaptive neural network approach, evolving alongside emerging threats, to detect unknown cyber-attacks effectively. The challenge lies in ensuring that models do not compromise detection performance for imbalanced data, which is crucial for low-resource and distributed environments.

Recent studies emphasize real-time detection and adaptability to changing threat landscapes. \citet{wu2024current} highlighted the need for efficient NIDS frameworks capable of handling zero-day attacks while maintaining low false positives through adversarial training and adaptive architectures. Similarly, \citet{dang2024adaptive} introduced a communication-efficient method, DaZoo, for distributed IoT environments, which enhances adaptability by reducing communication overhead in real-time applications. The challenge lies in balancing adaptability without compromising accuracy or introducing excessive computational complexity.

Several studies aim to integrate innovative methodologies to improve the detection of unknown attacks. For \citet{al2020unknown} proposed a hybrid approach for handling Type-A and Type-B unknown attacks using shallow and deep neural networks. Their approach highlighted the need for innovative models that combine lightweight classifiers for improved detection accuracy in IoT environments. However, defining consistent and systematic methodologies for unknown attack detection remains an ongoing challenge.

\section{Methodology}
\begin{figure*}[ht]
\centering
\includegraphics[width=\textwidth,frame]{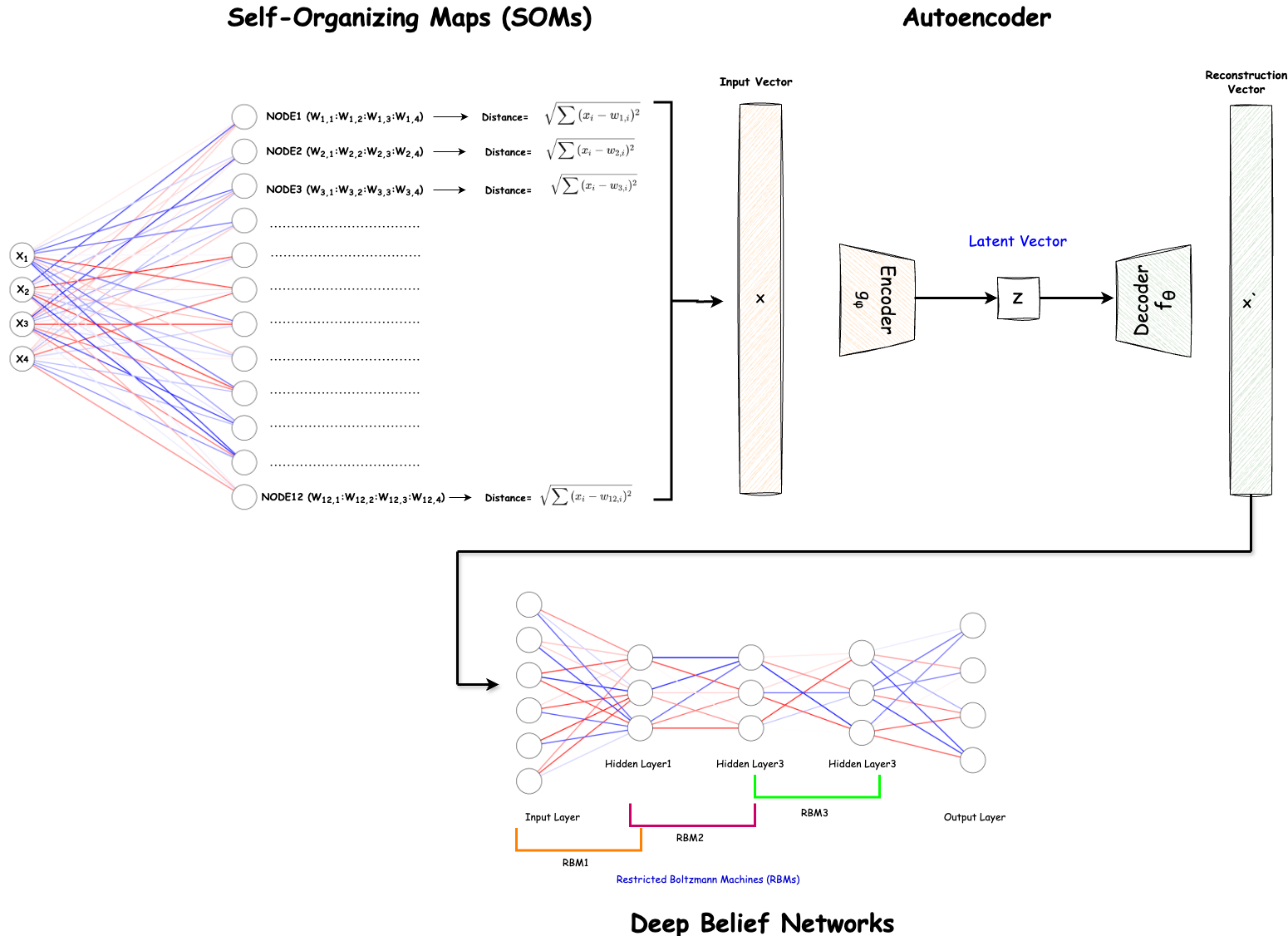}
\caption{Integration of SOMs, DBNs, and Autoencoders for Model Framework.}
\label{fig:Framwork}
\end{figure*}

To effectively address the complexities of cyber-attack detection in IoT networks, the proposed methodology integrates several machine learning techniques, including Self-Organizing Maps (SOMs), Deep Belief Networks (DBNs), and Autoencoders. This approach leverages the strengths of each technique to enhance the accuracy and robustness of attack detection systems. The methodology begins with the application of SOMs for unsupervised clustering and anomaly detection, followed by DBNs to capture hierarchical representations of the data. Autoencoders are then employed to perform dimensionality reduction and feature extraction, facilitating improved model performance, see Figure  \ref{fig:Framwork}. Additionally, an advanced optimization algorithm is incorporated to fine-tune the hyperparameters of these models, ensuring optimal performance in real-world scenarios.

\subsection{Self-Organizing Maps (SOMs)}
Self-Organizing Maps (SOMs) are a type of artificial neural network that utilize unsupervised learning to transform high-dimensional data into a lower-dimensional representation while preserving the topological relationships of the input space. This capability makes SOMs a powerful tool for data visualization and clustering tasks. The SOM topology comprises two fully connected layers: an input layer representing the data and an output layer arranged in a lattice of nodes. Each node in the output layer represents a specific region of the input space, with a neighborhood function maintaining the spatial relationships among nodes \citet{kohonen1997exploration}. The SOM learning process begins with random initialization of the weights. An input vector is then randomly selected and presented to the lattice. Each node computes its distance from the input vector using a distance metric, such as the Euclidean distance. The node with the smallest distance is identified as the Best Matching Unit (BMU). A neighborhood radius is then defined around the BMU, and the weights of the neighboring nodes are updated to more closely resemble the input vector. This competitive learning process ensures that the SOM learns to map similar inputs to adjacent regions in the output lattice.

\textbf{Role:} In the hybrid framework, SOMs act as a pre-processing/feature extraction component. They help in visualizing and grouping the input data in such a way that it makes it easier for the subsequent layers (e.g., DBNs and AEs) to perform classification or further analysis.

Self-Organizing Maps (SOMs) are employed to cluster IoT traffic data into distinct categories, allowing for effective anomaly detection. The SOM algorithm involves the following steps:

\subsubsection{Initialization}
Initialize the weight vectors $w_i(0)$ for each neuron $i$ in the map randomly within the input space.

\subsubsection{Training Process}
For each input vector $x(t)$, identify the Best Matching Unit (BMU) as follows:
\begin{equation}
\text{BMU} = \arg \min_i \| x(t) - w_i(t) \|
\end{equation}

Update the weights of the BMU and its neighbors using:
\begin{equation}
w_i(t+1) = w_i(t) + \eta(t) \cdot h_{i,\text{BMU}}(t) \cdot (x(t) - w_i(t))
\end{equation}

where $h_{i,\text{BMU}}(t)$ is the neighborhood function:
\begin{equation}
h_{i,\text{BMU}}(t) = \exp \left( -\frac{\|x(t) - w_i(t)\|^2}{2 \sigma^2(t)} \right)
\end{equation}

The learning rate $\eta(t)$ and neighborhood width $\sigma(t)$ decay over time:
\begin{equation}
\eta(t) = \eta_0 \cdot \exp \left( -\frac{t}{\tau_\eta} \right)
\end{equation}

\begin{equation}
\sigma(t) = \sigma_0 \cdot \exp \left( -\frac{t}{\tau_\sigma} \right)
\end{equation}

where $\tau_\eta$ and $\tau_\sigma$ are decay parameters.

\subsection{Deep Belief Networks (DBNs)}

Deep Belief Networks (DBNs) are utilized to learn hierarchical representations of IoT traffic data. DBNs are built from multiple Restricted Boltzmann Machines (RBMs) \cite{hinton2012practical} stacked together, allowing them to capture complex patterns and dependencies across different layers of the data. This architecture enables DBNs to effectively model the structure of high-dimensional IoT traffic, making them particularly useful for tasks such as anomaly detection, feature extraction, and predictive modeling. By learning a deep hierarchical structure, DBNs can better understand the underlying relationships and features present in the data, leading to more robust and accurate analyses. Furthermore, the probabilistic nature of RBMs within DBNs allows the incorporation of uncertainty and probabilistic reasoning into the decision-making process.

\textbf{Role:} In the hybrid framework, DBNs are used for further processing the features extracted by SOMs. By fine-tuning the representations learned by SOMs, DBNs help in learning hierarchical feature representations, making them more suitable for classification tasks.
\subsubsection{Restricted Boltzmann Machines (RBMs)}
The energy function of an RBM is:
\begin{equation}
E(x, h) = - \sum_{i} b_i x_i - \sum_{j} c_j h_j - \sum_{i,j} w_{ij} x_i h_j
\end{equation}

The joint probability distribution over visible and hidden units is:
\begin{equation}
P(x, h) = \frac{1}{Z} \exp \left( -E(x, h) \right)
\end{equation}

where $Z$ is the partition function. The learning algorithm uses Contrastive Divergence to approximate the gradient:
\begin{equation}
\nabla_\theta \mathcal{L} = \mathbb{E}_{\text{data}}[x h^T] - \mathbb{E}_{\text{model}}[x h^T]
\end{equation}

where $\mathbb{E}_{\text{data}}$ and $\mathbb{E}_{\text{model}}$ are the expectations with respect to the data distribution and the model distribution, respectively.

\subsubsection{Training DBNs}
To train DBNs, each RBM is trained in a layer-wise fashion. The top RBM is trained using the features extracted by the previous layers. The overall objective function for fine-tuning the DBN is:
\begin{equation}
\mathcal{L}(\theta) = \sum_{n=1}^{N} \log P(x^{(n)}; \theta)
\end{equation}

where $\theta$ includes all parameters of the DBN.

\subsection{Autoencoders}

Autoencoders are employed to perform dimensionality reduction and feature extraction. The architecture consists of an encoder and a decoder, where the encoder maps the input data into a lower-dimensional latent space, capturing the most relevant features while discarding redundant or irrelevant information. The decoder then reconstructs the original data from the reduced representation, ensuring that the key characteristics are preserved. This process helps to reduce noise and improve the efficiency of downstream tasks such as classification, clustering, and anomaly detection. Additionally, autoencoders can be fine-tuned to adapt to specific domains or datasets, making them highly versatile for various applications in data preprocessing and feature engineering.

\textbf{Role:} In the hybrid framework, AEs are used to learn a compressed representation of the input data and reduce noise. This makes the data more manageable for the DBNs, improving the classification task by ensuring only the most important features are passed through.

\subsubsection{Encoder and Decoder}
The encoder function is:
\begin{equation}
z = f_{\text{enc}}(x; \theta_{\text{enc}})
\end{equation}

The decoder function is:
\begin{equation}
\hat{x} = f_{\text{dec}}(z; \theta_{\text{dec}})
\end{equation}

where $\theta_{\text{enc}}$ and $\theta_{\text{dec}}$ are the parameters of the encoder and decoder, respectively. The reconstruction loss is:
\begin{equation}
\mathcal{L}_{\text{rec}} = \frac{1}{N} \sum_{n=1}^{N} \| x^{(n)} - \hat{x}^{(n)} \|^2
\end{equation}

\subsubsection{Regularization}
To enhance the performance of Autoencoders, we apply regularization techniques:
\begin{equation}
\mathcal{L}_{\text{reg}} = \mathcal{L}_{\text{rec}} + \lambda \| \theta_{\text{enc}} \|_2^2
\end{equation}

where $\lambda$ is the regularization parameter.

\subsection{Novel Optimization Algorithm}

The proposed optimization algorithm integrates Particle Swarm Optimization (PSO) with the other techniques. It aims to optimize the hyperparameters for SOMs, DBNs, and Autoencoders simultaneously.

\subsubsection{Initialization}
Initialize particles with hyperparameters for DBN, SOM, and Autoencoder. Each particle represents a potential solution in the hyperparameter space.

\subsubsection{Particle Update Rules}
Update the velocity $\vec{v}_i$ and position $\vec{x}_i$ of each particle:
\begin{equation}
\vec{v}_i(t+1) = \omega \vec{v}_i(t) + c_1 r_1 (\vec{p}_i - \vec{x}_i(t)) + c_2 r_2 (\vec{g} - \vec{x}_i(t))
\end{equation}

\begin{equation}
\vec{x}_i(t+1) = \vec{x}_i(t) + \vec{v}_i(t+1)
\end{equation}

where $\omega$ is the inertia weight, $c_1$ and $c_2$ are cognitive and social coefficients, and $r_1$ and $r_2$ are random values.

\subsubsection{Hyperparameter Optimization}
For each particle, train the DBN, SOM, and Autoencoder using the hyperparameters represented by the particle. Compute the cost function:
\begin{equation}
\mathcal{L}(\theta) = \alpha \mathcal{L}_{\text{rec}} + \beta \mathcal{L}_{\text{clus}} + \gamma \mathcal{L}_{\text{hier}}
\end{equation}

where $\mathcal{L}_{\text{clus}}$ is the clustering loss from SOMs, and $\mathcal{L}_{\text{hier}}$ is the hierarchical loss from DBNs.

\subsubsection{Adaptive Learning Rate Adjustments}
Adjust the learning rates for DBNs and Autoencoders based on performance:
\begin{equation}
\eta_{t} = \frac{\eta_{0}}{1 + \alpha t}
\end{equation}

where $\eta_{0}$ is the initial learning rate and $\alpha$ is the decay parameter.

\subsection{Interaction Between SOMs, DBNs, and AEs}
\noindent\textbf{Preprocessing Stage (SOMs):} The input data is first processed by the SOM to perform clustering and dimensionality reduction. SOMs help in grouping similar data points together, making it easier for the model to focus on the relevant parts of the data.

\noindent\textbf{Feature Refinement Stage (AEs):} After SOMs, Autoencoders refine these features by compressing the data into a more compact form while removing any noise. This step ensures that the data passed to DBNs has a more focused and optimized representation.

\noindent\textbf{Classification Stage (DBNs):} Finally, the processed and optimized data is passed through DBNs, where hierarchical and abstract representations are learned. The output layer of the DBN is typically designed for classification, mapping the learned features to the desired class labels.

\subsection{Hyperparameter Tuning with Particle Swarm Optimization (PSO)}

In the context of this hybrid framework, PSO plays a crucial role in hyperparameter tuning:

\noindent\textbf{Application in SOMs:} PSO optimizes the SOM parameters, such as the size of the grid and the learning rate. For example, the neighborhood size is set to 10, and the learning rate is set to 0.05. By exploring the search space efficiently, PSO helps find the optimal SOM configuration that best clusters the data, thus enhancing its effectiveness in data preprocessing and feature extraction.

\noindent\textbf{Application in DBNs:} PSO is used to optimize hyperparameters like the number of layers, the number of hidden units in each layer, and the learning rate. For instance, a typical configuration involves 3 layers, with 256 units per layer and a learning rate of 0.005. PSO searches for the best combination of these parameters to improve the feature learning capacity of the DBN, which in turn improves the classification accuracy by better capturing complex patterns in the data.

\noindent\textbf{Application in AEs:} In the case of Autoencoders, PSO tunes the architecture of the AE, such as the number of neurons in the encoder and decoder layers and the activation functions used. For instance, the latent space dimension is set to 128, and the reconstruction error threshold is 0.01. Additionally, PSO optimizes regularization parameters, such as the weight decay factor, to avoid overfitting and underfitting, ensuring the model generalizes well to unseen data.

\noindent\textbf{Overall Influence:} PSO improves the overall performance of the hybrid framework by efficiently exploring the hyperparameter search space for each component model. By optimizing these hyperparameters, PSO ensures that SOMs, DBNs, and AEs are operating in the most effective configuration for the specific data and classification task at hand. This not only enhances individual model performance but also improves the synergy between the models in the hybrid framework, leading to better classification outcomes.

\subsection{Algorithm for Advanced Cyber-Attack Detection}
This section outlines a hybrid algorithm for detecting cyber-attacks in IoT networks, integrating Self-Organizing Maps (SOMs), Deep Belief Networks (DBNs), and Autoencoders, optimized via Particle Swarm Optimization (PSO). The approach enhances detection accuracy and adaptability by combining anomaly detection, feature learning, and reconstruction-based methods.\\

\textbf{Algorithm Workflow}
\begin{enumerate}[]
    \item \textbf{Input:} Dataset $D = \{X, Y\}$, initial hyperparameters $H$, and PSO parameters.
    \item \textbf{Preprocessing:} Normalize and reduce feature dimensions in $X$.
    \item \textbf{Anomaly Detection:} Train SOMs to detect anomalous patterns in $X$.
    \item \textbf{Feature Learning:} Use DBNs to extract hierarchical features $F_{\text{DBN}}$.
    \item \textbf{Reconstruction-Based Detection:} Train Autoencoders on $F_{\text{DBN}}$ and calculate reconstruction errors.
    \item \textbf{Optimization:} Use PSO to tune hyperparameters for SOMs, DBNs, and Autoencoders, maximizing performance metrics.
    \item \textbf{Output:} Classify attacks or anomalies with optimized models.
\end{enumerate}

\begin{algorithm*}
\caption{Advanced Cyber-Attack Detection Framework}
\begin{algorithmic}[1]
\STATE \textbf{Input:} Network traffic data $X$, Labels $Y$, Hyperparameters for SOM, DBN, and Autoencoders.
\STATE \textbf{Output:} Optimized attack detection model, Detection results.

\STATE \textbf{Step 1: Data Preprocessing}
\STATE Normalize data:
\[
X_{\text{norm}} = \frac{X - \mu}{\sigma}
\]
\STATE Split into train and test sets: $(X_{\text{train}}, Y_{\text{train}})$, $(X_{\text{test}}, Y_{\text{test}})$.

\STATE \textbf{Step 2: Dimensionality Reduction with Autoencoders}
\STATE Train Autoencoder:
\[
\mathcal{L}_{\text{rec}} = \frac{1}{N} \sum_{i=1}^{N} \| X_{\text{train}} - \hat{X}_{\text{train}} \|^2
\]
\STATE Extract compressed features $X_{\text{compressed}}$.

\STATE \textbf{Step 3: Anomaly Detection with SOM}
\STATE Assign data to Best Matching Unit (BMU):
\[
\text{BMU} = \arg \min_i \| X_{\text{compressed}} - w_i \|
\]
\STATE Update BMU weights:
\[
w_i(t+1) = w_i(t) + \eta(t) h_{i,\text{BMU}}(t) \left( X_{\text{compressed}}(t) - w_i(t) \right)
\]
\STATE Detect anomalies as outliers.

\STATE \textbf{Step 4: Classification with DBN}
\STATE Train Deep Belief Network:
\[
\mathcal{L}_{\text{cls}} = -\frac{1}{N} \sum_{i=1}^{N} Y_i \log \hat{Y}_i
\]
\STATE Fine-tune DBN using backpropagation.

\STATE \textbf{Step 5: Hyperparameter Optimization with PSO}
\STATE Update particle velocities:
\[
\vec{v}_i(t+1) = \omega \vec{v}_i(t) + c_1 r_1 (\vec{p}_i - \vec{x}_i(t)) + c_2 r_2 (\vec{g} - \vec{x}_i(t))
\]
\STATE Update positions:
\[
\vec{x}_i(t+1) = \vec{x}_i(t) + \vec{v}_i(t+1)
\]
\STATE Evaluate fitness and select best hyperparameters.

\STATE \textbf{Step 6: Integration of Anomaly Detection and Classification}
\STATE Combine SOM anomaly detection and DBN classification results.

\STATE \textbf{Step 7: Model Evaluation}
\STATE Evaluate using accuracy, precision, recall, and F1-score.

\end{algorithmic}
\end{algorithm*}

\section{Datasets}

This section discusses the datasets used in our experiments, including the data preprocessing techniques, feature selection methods, and a detailed breakdown of the features for each dataset. We employ two well-known network intrusion datasets: \textbf{NSL-KDD} and \textbf{UNSW-NB15}, and \textbf{CICIoT2023} all of which offer a variety of network traffic patterns to evaluate our models' performance.

\subsection{NSL-KDD Dataset}
The NSL-KDD dataset \cite{dhanabal2015study} is a revised version of the KDD99 dataset, designed to eliminate redundancy while preserving the diversity of attack types. It comprises 41 features as shown in Table \ref{tab:nslkdd-features} and a label indicating whether the traffic is normal or one of four attack types: \textit{Denial of Service (DoS)}, \textit{Probe}, \textit{User-to-Root (U2R)}, and \textit{Remote-to-Local (R2L)}. The dataset is divided into training and testing sets with a diverse mix of attacks. In this experiment, we focus primarily on DoS attacks and normal traffic.

\subsection{UNSW-NB15 Dataset}
The UNSW-NB15 dataset \cite{moustafa2015unsw} created by the Australian Centre for Cyber Security (ACCS) to simulate real-world attack scenarios. This dataset contains 44 features as shown in Table \ref{tab:unsw-features}, divided into 39 numerical values and 4 categorical values, with a label indicating whether the traffic is normal or one of the nine attack types: \textit{Generic}, \textit{Reconnaissance}, \textit{DoS}, \textit{Exploits}, \textit{Analysis}, \textit{Worms}, \textit{Shellcode}, \textit{Backdoor}, and \textit{Fuzzers}. For our experiments, we focused on the \textit{DoS}, \textit{Exploits}, \textit{Generic}, and \textit{Fuzzers} attack types.

\subsection{CICIoT2023 Dataset}
The CICIoT2023 dataset, developed by the Canadian Institute for Cybersecurity \cite{neto2023ciciot2023}, is designed to support research on secure analysis applications for large-scale IoT (Internet of Things) attacks. It includes both normal traffic and 33 types of attack traffic, organized into seven categories: DDoS, DoS, Reconnaissance, Web-based, Brute Force, Spoofing, and Mirai. Table \ref{tab:CICIoT2023DatasetFeatures} provides a detailed overview of the dataset. All attack traffic originates from malicious IoT devices targeting other IoT devices, making it an ideal resource for studies on intrusion detection and related IoT security challenges.

\subsection{Data Preprocessing}
Data preprocessing is crucial for improving the model's accuracy and reducing computational complexity. First, redundant and irrelevant features are removed from the datasets. Categorical features are encoded into numerical values using \textit{Label Encoding} and \textit{One-Hot Encoding}. Finally, \textit{Min-Max Normalization} is applied to scale the feature values between 0 and 1, following the equation:

\begin{equation}
    x' = \frac{x - \min(x)}{\max(x) - \min(x)}
\end{equation}

where $x$ is the original value, and $\min(x)$ and $\max(x)$ are the minimum and maximum values of the feature, respectively. This transformation enhances model performance by normalizing the data.

\subsection{Feature Selection}
Feature selection reduces the dimensionality of the dataset, enhancing model efficiency by removing irrelevant or redundant features. We utilized three feature selection methods:

\subsubsection{Correlation-Based Filter Method}
This method evaluates the correlation between features using a correlation matrix. Highly correlated features, determined by a threshold $\theta$, are removed to avoid multicollinearity. The correlation score $\rho$ is computed as:

\begin{equation}
    \rho_{ij} = \frac{\text{cov}(X_i, X_j)}{\sigma_{X_i} \sigma_{X_j}}
\end{equation}

where $\text{cov}(X_i, X_j)$ is the covariance between features $X_i$ and $X_j$, and $\sigma_{X_i}$ and $\sigma_{X_j}$ are their standard deviations. If $\rho_{ij} > \theta$, one of the features is discarded.

\subsubsection{Wrapper Methods}
Wrapper methods evaluate feature subsets by iteratively selecting features that improve model performance. We applied both \textit{Forward Selection} and \textit{Backward Elimination}. Forward selection starts with an empty feature set, adding features that improve performance, while backward elimination starts with all features, removing those that decrease performance.

\subsubsection{Embedded Methods}
Embedded methods integrate feature selection within the model training process. We employed \textit{LASSO Regularization} for this purpose, which penalizes the magnitude of coefficients to zero for irrelevant features. The LASSO objective function is:

\begin{equation}
    \min \left( \frac{1}{2n} \sum_{i=1}^{n} \left( y_i - \hat{y}_i \right)^2 + \lambda \sum_{j=1}^{p} | \beta_j | \right)
\end{equation}

where $\lambda$ is the regularization parameter, $y_i$ is the true label, $\hat{y}_i$ is the predicted label, and $\beta_j$ is the coefficient of feature $j$.

\subsection{Feature Tables}

The features of both datasets are detailed in Tables \ref{tab:nslkdd-features} and \ref{tab:unsw-features}.

\begin{table*}[h]
\caption{NSL-KDD dataset features.}
\label{tab:nslkdd-features}
\begin{center}
\begin{tabular}{ccc ccc}
\toprule
S.No & Feature name & S.No & Feature name & S.No & Feature name \\
\midrule
1 & protocol\_type & 15 & num\_shells & 29 & srv\_rerror\_rate \\
2 & src\_bytes & 16 & num\_access\_files & 30 & root\_shell \\
3 & rerror\_rate & 17 & serror\_rate & 31 & dst\_host\_diff\_srv\_rate \\
4 & is\_guest\_login & 18 & dst\_host\_serror\_rate & 32 & num\_root \\
5 & srv\_serror\_rate & 19 & duration & 33 & dst\_host\_same\_src\_port\_rate \\
6 & flag & 20 & srv\_diff\_host\_rate & 34 & srv\_diff\_host\_rate \\
7 & service & 21 & srv\_count & 35 & dst\_host\_srv\_serror\_rate \\
8 & num\_failed\_logins & 22 & wrong\_fragment & 36 & num\_file\_creations \\
9 & dst\_host\_count & 23 & dst\_bytes & 37 & dst\_host\_srv\_rerror\_rate \\
10 & num\_compromised & 24 & land & 38 & num\_outbound\_cmds \\
11 & dst\_host\_same\_srv\_rate & 25 & hot & 39 & dst\_host\_srv\_count \\
12 & su\_attempted & 26 & urgent & 40 & same\_srv\_rate \\
13 & dst\_host\_srv\_diff\_host\_rate & 27 & num\_root & 41 & diff\_srv\_rate \\
14 & dst\_host\_rerror\_rate & 28 & logged\_in & 42 & class \\
\bottomrule
\end{tabular}
\end{center}
\end{table*}

\begin{table*}[h]
\caption{UNSW-NB15 dataset features.}
\label{tab:unsw-features}
\begin{center}
\begin{tabular}{ccc ccc}
\toprule
S.No & Feature name & S.No & Feature name & S.No & Feature name \\
\midrule
1 & Dur & 16 & djit & 31 & trans\_depth \\
2 & dpkts & 17 & ct\_dst\_src\_ltm & 32 & response\_body\_len \\
3 & spkts & 18 & ct\_ftp\_cmd & 33 & ct\_srv\_src \\
4 & dbytes & 19 & ct\_src\_ltm & 34 & is\_ftp\_login \\
5 & sbytes & 20 & is\_sm\_ips\_ports & 35 & ct\_dst\_ltm \\
6 & rate & 21 & swin & 36 & ct\_dst\_sport\_ltm \\
7 & Sttl & 22 & attack\_cat & 37 & ct\_src\_dport\_ltm \\
8 & dttl & 23 & Stcpb & 38 & ct\_state\_ttl \\
9 & sload & 24 & dtcpb & 39 & ct\_flw\_http\_mthd \\
10 & dload & 25 & dwin & 40 & ct\_srv\_dst \\
11 & sloss & 26 & tcprtt & 41 & proto \\
12 & dloss & 27 & synack & 42 & service \\
13 & sinpkt & 28 & ackdat & 43 & state \\
14 & dinpkt & 29 & smean & 44 & label \\
15 & sjit & 30 & dmean &  &  \\
\bottomrule
\end{tabular}
\end{center}
\end{table*}

\begin{table*}[h]
\caption{CICIoT2023 dataset features.}
\label{tab:CICIoT2023DatasetFeatures} 
\begin{tabular}{@{}lll@{}}
\toprule
S.No & Feature name & Description                                                                                                                                                               \\ \midrule
1             & ts                    & Timestamp                                                                                                                                                                          \\
2             & flow duration         & Duration of the packet’s flow                                                                                                                                                      \\
3             & Header Length         & Header Length                                                                                                                                                                      \\
4             & Protocol Type         & IP, UDP, TCP, IGMP, ICMP, Unknown (Integers)                                                                                                                                       \\
5             & Duration              & Time-to-Live (ttl)                                                                                                                                                                 \\
6             & Rate                  & Rate of packet transmission in a flow                                                                                                                                              \\
7             & Srate                 & Rate of outbound packets transmission in a flow                                                                                                                                    \\
8             & Drate                 & Rate of inbound packets transmission in a flow                                                                                                                                     \\
9             & fin flag number       & Fin flag value                                                                                                                                                                     \\
10            & syn flag number       & Syn flag value                                                                                                                                                                     \\
11            & rst flag number       & Rst flag value                                                                                                                                                                     \\
12            & psh flag numbe        & Psh flag value                                                                                                                                                                     \\
13            & ack flag number       & Ack flag value                                                                                                                                                                     \\
14            & ece flag numbe        & Ece flag value                                                                                                                                                                     \\
15            & cwr flag number       & Cwr flag value                                                                                                                                                                     \\
16            & ack count             & Number of packets with ack flag set in the same flow                                                                                                                               \\
17            & syn count             & Number of packets with syn flag set in the same flow                                                                                                                               \\
18            & fin count             & Number of packets with fin flag set in the same flow                                                                                                                               \\
19            & urg coun              & Number of packets with urg flag set in the same flow                                                                                                                               \\
20            & rst count             & Number of packets with rst flag set in the same flow                                                                                                                               \\
21            & HTTP                  & Indicates if the application layer protocol is HTTP                                                                                                                                \\
22            & HTTPS                 & Indicates if the application layer protocol is HTTPS                                                                                                                               \\
23            & DNS                   & Indicates if the application layer protocol is DNS                                                                                                                                 \\
24            & Telnet                & Indicates if the application layer protocol is Telnet                                                                                                                              \\
25            & SMTP                  & Indicates if the application layer protocol is SMTP                                                                                                                                \\
26            & SSH                   & Indicates if the application layer protocol is SSH                                                                                                                                 \\
27            & IRC                   & Indicates if the application layer protocol is IRC                                                                                                                                 \\
28            & TCP                   & Indicates if the transport layer protocol is TCP                                                                                                                                   \\
29            & UDP                   & Indicates if the transport layer protocol is UDP                                                                                                                                   \\
30            & DHCP                  & Indicates if the application layer protocol is DHCP                                                                                                                                \\
31            & ARP                   & Indicates if the link layer protocol is ARP                                                                                                                                        \\
32            & ICMP                  & Indicates if the network layer protocol is ICMP                                                                                                                                    \\
33            & IPv                   & Indicates if the network layer protocol is IP                                                                                                                                      \\
34            & LLC                   & Indicates if the link layer protocol is LLC                                                                                                                                        \\
35            & Tot sum               & Summation of packets lengths in flow                                                                                                                                               \\
36            & Min                   & Minimum packet length in the flow                                                                                                                                                  \\
37            & Max                   & Maximumpacket length in the flow                                                                                                                                                   \\
38            & Avg                   & Average packet length in the flow                                                                                                                                                  \\
39            & Std                   & Standard deviation of packet length in the flow                                                                                                                                    \\
40            & Tot size              & Packet’s length                                                                                                                                                                    \\
41            & IAT                   & The time difference with the previous packet                                                                                                                                       \\
42            & Number                & The number of packets in the flow                                                                                                                                                  \\
43            & Magnitude             & \begin{tabular}[c]{@{}l@{}}(Average of the lengths of incoming packets in the flow +\\ average of the lengths of outgoing packets in the flow)$^{0.5}$\end{tabular}   \\
44            & Radius                & \begin{tabular}[c]{@{}l@{}}(Variance of the lengths of incoming packets in the flow +\\ variance of the lengths of outgoing packets in the flow)$^{0.5}$\end{tabular} \\
45            & Covariance            & Covariance of the lengths of incoming and outgoing packets                                                                                                                         \\
46            & Variance              & \begin{tabular}[c]{@{}l@{}}Variance of the lengths of incoming packets in the flow/\\ variance of the lengths of outgoing packets in the flow\end{tabular}                         \\
47            & Weight                & Number of incoming packets $\times$ Number of outgoing packets                                                                                                        \\ \bottomrule
\end{tabular}
\end{table*}

\section{Experimental Setup}

The proposed methodology for cyber-attack detection in IoT networks is evaluated using a comprehensive experimental setup. This section details the data, evaluation metrics, and configurations employed to assess the performance of Self-Organizing Maps (SOMs), Deep Belief Networks (DBNs), and Autoencoders. The experiments are conducted in a controlled environment, utilizing both simulated and real-world IoT traffic data.

\subsection{Model Configurations}

The models are configured as follows:

\begin{itemize} \item \textbf{Self-Organizing Maps (SOMs)}: A grid size of $10 \times 10$, with an initial learning rate of 0.1 and a decay rate of 0.01. The neighborhood function utilizes a Gaussian kernel with an initial radius of 3, decaying over time. \item \textbf{Deep Belief Networks (DBNs)}: Comprising three layers of Restricted Boltzmann Machines (RBMs) with 128, 64, and 32 hidden units. The learning rate is set to 0.01, with the contrastive divergence algorithm employed for training. \item \textbf{Autoencoders}: Featuring an encoder with two hidden layers (128 and 64 units) and a symmetric decoder structure. The activation function is ReLU, and the model is optimized using Adam with a learning rate of 0.001. \end{itemize}

Additionally, the following parameters are subject to Particle Swarm Optimization (PSO) for optimization: Learning rate, Number of hidden units (for DBNs and Autoencoders), Neighborhood size (for SOMs), and Activation functions.

\subsection{Optimization Strategy}

A Particle Swarm Optimization (PSO) algorithm \cite{wang2018particle} optimizes the hyperparameters of SOMs, DBNs, and Autoencoders. The search space includes:

\begin{itemize} \item Learning rate: [0.0001, 0.01] \item Number of hidden units: [32, 128] \item Neighborhood size (for SOMs): [1, 5] \item Activation functions: ReLU, Sigmoid \end{itemize}

The PSO algorithm runs for 100 iterations with 30 particles in the swarm, with the fitness function defined as a weighted sum of accuracy, precision, and recall.

\subsection{Computational Environment}

Experiments are conducted on a system with:

\begin{itemize} \item Processor: Intel Core i7-10750H CPU @ 2.60GHz \item RAM: 32GB DDR4 \item GPU: NVIDIA RTX 2080 (8GB) \item Software: TensorFlow, Keras, and SciPy for model training and optimization. \end{itemize}

Experiments are repeated five times with different random seeds, and average results are reported.

\section{Results}

The proposed model evaluated on two widely used cybersecurity datasets: NSL-KDD, UNSW-NB15 and CICIoT2023. The results demonstrate excellent performance in detecting both known and unknown attack types. Additionally, the model's robustness in identifying unknown attack instances highlights its effectiveness in addressing real-world security challenges.

The tables \ref{tab:NSL-KDDDatasetFeatures}, \ref{tab:NSL-UNSW-NB15Features}, and \ref{tab:CICIoT2023datasetattack} summarize the attack types found in three datasets, and we compare detection rates, focusing on unknown attack types.

\begin{table}[H]
\centering
\caption{Attack types found in the NSL-KDD dataset.}
\label{tab:NSL-KDDDatasetFeatures} 

\begin{tabular}{ll}
\toprule
Attack Category & Description \\ 
\midrule
DoS (Denial of Service)  & Exhaust system resources \\ 
R2L (Remote to Local)    & Unauthorized local access \\ 
U2R (User to Root)       & Unauthorized root access \\ 
Probe                    & Scan for vulnerabilities \\ 
Normal                   & Regular traffic \\ 
\bottomrule
\end{tabular}
\end{table}

\begin{table}[H]
\centering
\caption{Attack types found in the UNSW-NB15 dataset.}
\label{tab:NSL-UNSW-NB15Features} 
\begin{tabular}{ll}
\toprule
Attack Category & Description \\ 
\midrule
Exploits                 & Malicious payload injection \\ 
Fuzzers                  & Random input to crash systems \\ 
Backdoor                 & Unauthorized system access \\ 
DoS                      & Denial of Service attacks \\ 
Worms                    & Self-replicating malware \\ 
Analysis                 & Vulnerability probing \\ 
Reconnaissance           & Gathering system data \\ 
Normal                   & Regular network traffic \\ 
\bottomrule
\end{tabular}
\end{table}

\begin{table}[H]
\centering
\caption{Number of rows for each attack and category in the CICIoT2023 dataset.}
\label{tab:CICIoT2023datasetattack} 
\begin{tabular}{llr}
\toprule
Category & Attack & Rows \\
\midrule
\multirow{1}{*}{Benign} 
& Benign Traffic & 1,098,195 \\
\midrule
\multirow{12}{*}{DDoS} 
 & ACK Fragmentation & 285,104 \\
 & UDP Flood & 5,412,287 \\
 & SlowLoris & 23,426 \\
 & ICMP Flood & 7,200,504 \\
 & RSTFIN Flood & 4,045,285 \\
 & PSHACK Flood & 4,094,755 \\
 & HTTP Flood & 28,790 \\
 & UDP Fragmentation & 286,925 \\
 & ICMP Fragmentation & 452,489 \\
 & TCP Flood & 4,497,667 \\
 & SYN Flood & 4,059,190 \\
 & SynonymousIP Flood & 3,598,138 \\
\midrule
\multirow{4}{*}{DoS} 
 & TCP Flood & 2,671,445 \\
 & HTTP Flood & 71,864 \\
 & SYN Flood & 2,028,834 \\
 & UDP Flood & 3,318,595 \\
\midrule
\multirow{5}{*}{Recon} 
 & Ping Sweep & 2,262 \\
 & OS Scan & 98,259 \\
 & Vulnerability Scan & 37,382 \\
 & Port Scan & 82,284 \\
 & Host Discovery & 134,378 \\
\midrule
\multirow{6}{*}{Web-Based} 
 & SQL Injection & 5,245 \\
 & Command Injection & 5,409 \\
 & Backdoor Malware & 3,218 \\
 & Uploading Attack & 1,252 \\
 & XSS & 3,846 \\
 & Browser Hijacking & 5,859 \\
\midrule
Brute Force 
 & Dictionary Brute Force & 13,064 \\
\midrule
\multirow{2}{*}{Spoofing} 
 & ARP Spoofing & 307,593 \\
 & DNS Spoofing & 178,911 \\
\midrule
\multirow{3}{*}{Mirai} 
 & GREIP Flood & 751,682 \\
 & Greeth Flood & 991,866 \\
 & UDPPlain & 890,576 \\
\bottomrule
\end{tabular}
\end{table}

\begin{figure}[ht]
\centering
\includegraphics[width=8cm, height=6cm]{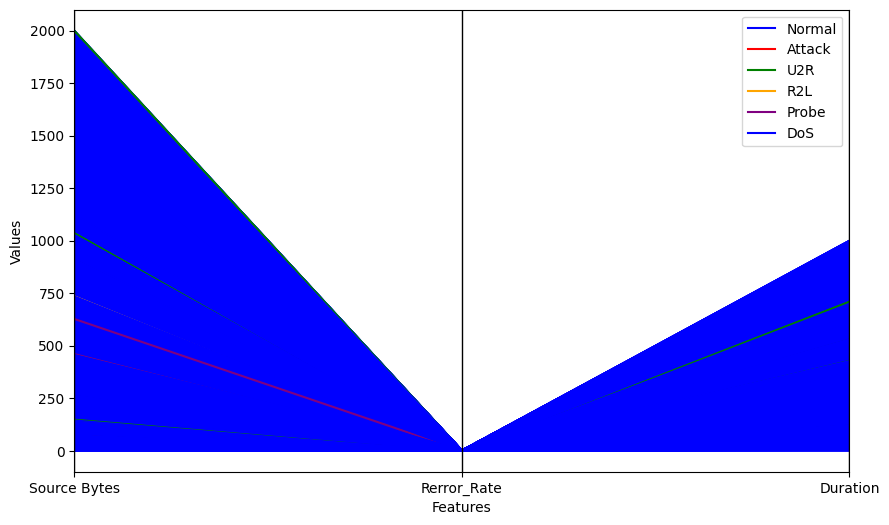}
\caption{Parallel Coordinates Plot showing the distribution of Source Bytes, Rerror Rate, and Duration across different class labels.}
\label{fig:Figure1}
\end{figure}

Figure \ref{fig:Figure1} illustrates a parallel coordinates plot comparing feature distributions for different attack classes in the NSL-KDD and UNSW-NB15 datasets. Each line represents a sample, with the features such as Source Bytes, Rerror Rate, and Duration plotted along the axes. The distinct colors signify different attack classes, allowing for visual differentiation of patterns across the datasets. This visualization aids in understanding how feature values vary between normal and attack instances, highlighting potential overlaps and unique characteristics that can inform anomaly detection algorithms.

\begin{figure}[ht]
\centering
\includegraphics[width=8cm, height=6cm]{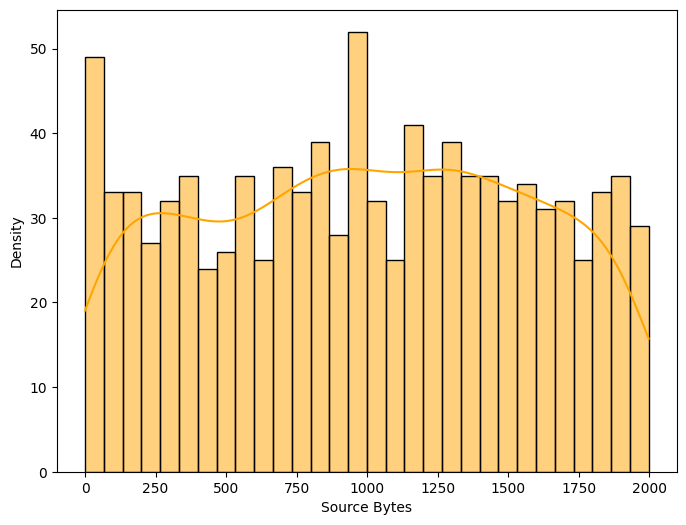}
\caption{Histogram of source bytes with KDE overlay, showing byte size distribution and insights into normal versus potentially malicious network traffic.}
\label{fig:Figure2}
\end{figure}

Figure \ref{fig:Figure2} displays the distribution of Source Bytes in the NSL-KDD dataset, with a kernel density estimate (KDE) overlayed to illustrate the probability density function. The x-axis represents the number of Source Bytes, while the y-axis shows the density of instances for each range of values. This plot indicates that most connections have relatively low Source Bytes, with a gradual decline in frequency as the Source Bytes increase. Understanding this distribution is crucial for analyzing the typical behavior of network traffic and detecting anomalies.

\begin{figure*}[ht]
\centering
\includegraphics[width=\textwidth]{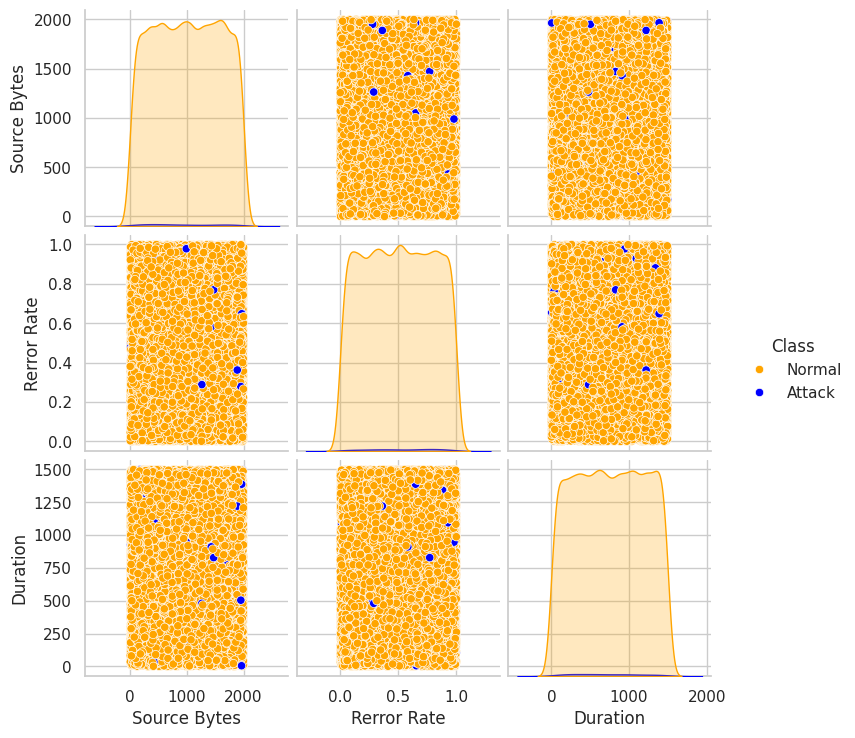}
\caption{Pairwise relationships and distributions of selected NSL-KDD features for normal and attack traffic.}

\label{fig:Figure3}
\end{figure*}
\begin{figure*}[ht]
\centering
\includegraphics[width=\textwidth]{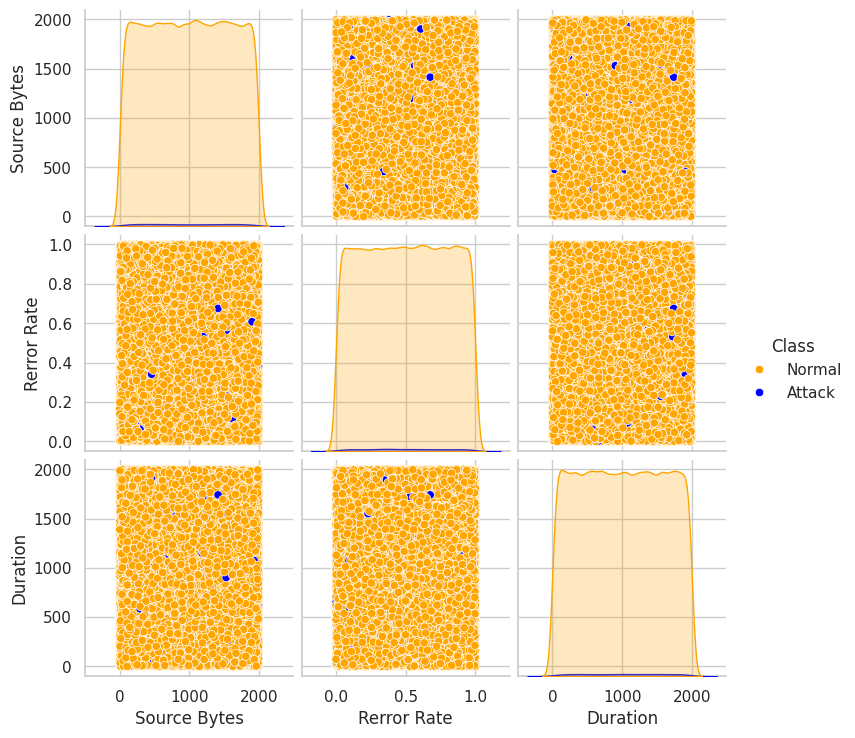}
\caption{Pairwise relationships and distributions of selected UNSW-NB15 features for normal and attack traffic.}
\label{fig:PairUNSWNB15}
\end{figure*}
\begin{figure*}[ht]
\centering
\includegraphics[width=\textwidth]{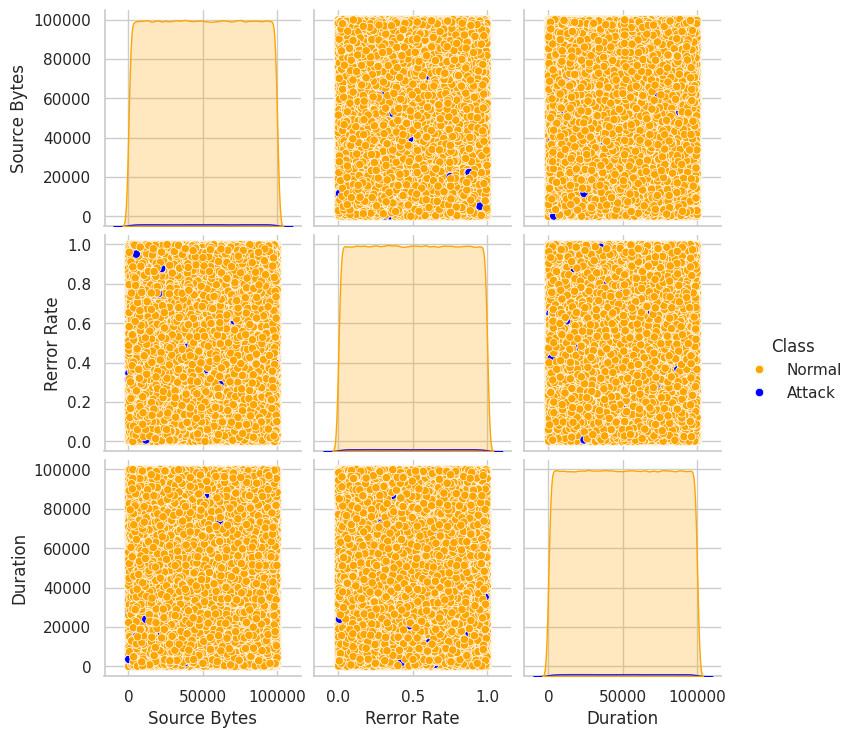}
\caption{Pairwise relationships and distributions of selected CICIoT2023 features for normal and attack traffic.}

\label{fig:PairCICIoT2023}
\end{figure*}

Figures \ref{fig:Figure3}, \ref{fig:PairUNSWNB15}, and \ref{fig:PairCICIoT2023} provide a comprehensive view of the relationships between the selected features: Source Bytes, Error Rate, Duration, and Class. Each scatter plot illustrates how two features correlate with each other, colored by their class labels. The diagonal displays the distribution of each feature, offering insights into their ranges and variations. This visualization facilitates the quick identification of how different features contribute to distinguishing between 'Normal' and 'Attack' classes, making it easier to detect patterns or overlaps in feature values.

\begin{figure*}[ht]
\centering
\includegraphics[width=\textwidth]{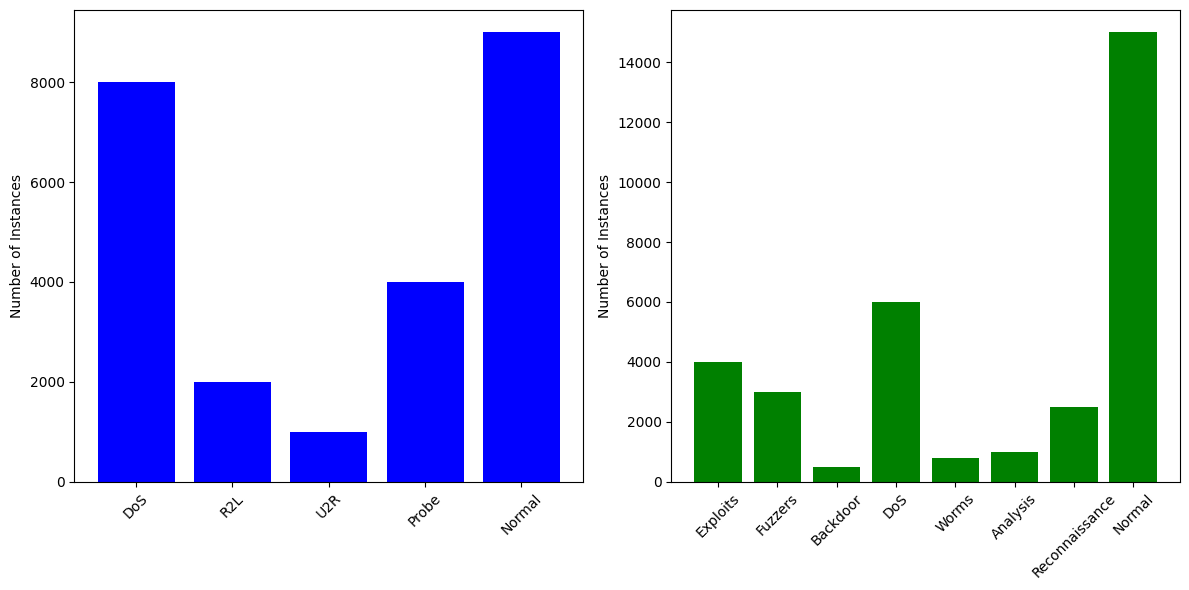}
\caption{Comparison of attack type distributions between NSL-KDD and UNSW-NB15 datasets using bar plots.}
\label{fig:Figure4}
\end{figure*}

Figure \ref{fig:Figure4} compares the number of instances for various attack types in the NSL-KDD and UNSW-NB15 datasets. The left plot shows the distribution of attack instances in NSL-KDD, with DoS attacks being the most prevalent, followed by Probe and R2L types. In contrast, the right plot for UNSW-NB15 highlights a more diverse set of attack types, with Exploits and DoS being the most common. This comparison not only provides insights into the attack landscape in each dataset but also helps in understanding the effectiveness of detection methods for different types of attacks.

\begin{figure*}[ht]
\centering
\includegraphics[width=\textwidth]{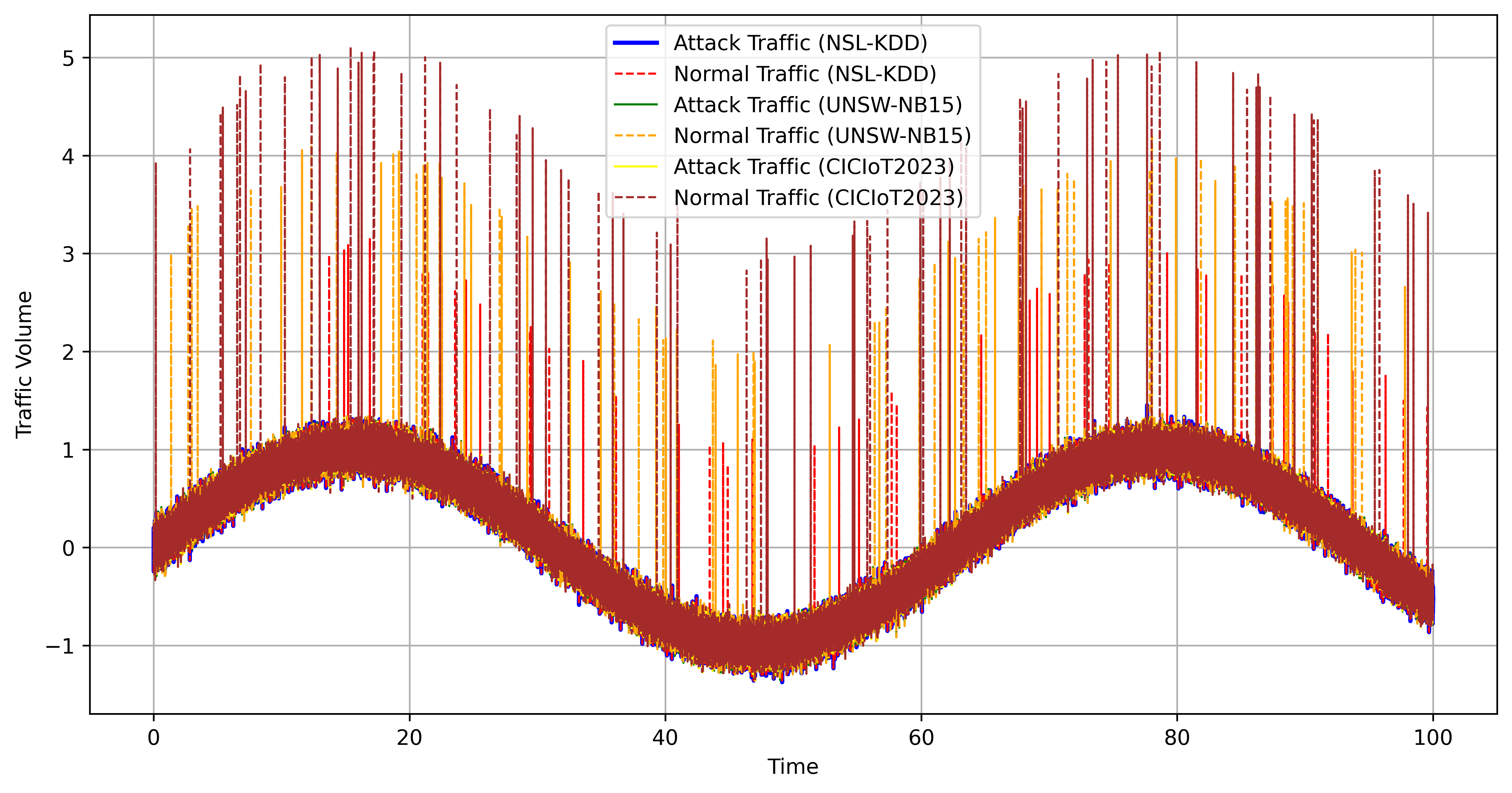}
\caption{IoT traffic analysis comparing normal and attack patterns.}
\label{fig:Figure5}
\end{figure*}

Figure \ref{fig:Figure5} illustrates the traffic patterns for both normal and attack scenarios across the NSL-KDD, UNSW-NB15, and CICIoT2023 datasets. The blue, green, and yellow lines represent normal traffic, while the red, orange, and brown dashed lines indicate attack traffic, with spikes corresponding to simulated attacks. This visualization highlights the effectiveness of the detection framework in distinguishing between regular and anomalous traffic patterns. The differences in attack magnitudes across datasets are evident, demonstrating the varying severity of attacks and the unique characteristics of each dataset. Also, the distinct attack behaviors, such as the higher magnitude spikes observed in CICIoT2023 compared to NSL-KDD and UNSW-NB15, further emphasize the need for tailored detection strategies to effectively handle different datasets.

\begin{figure}[ht]
\centering
\includegraphics[width=8cm, height=6cm]{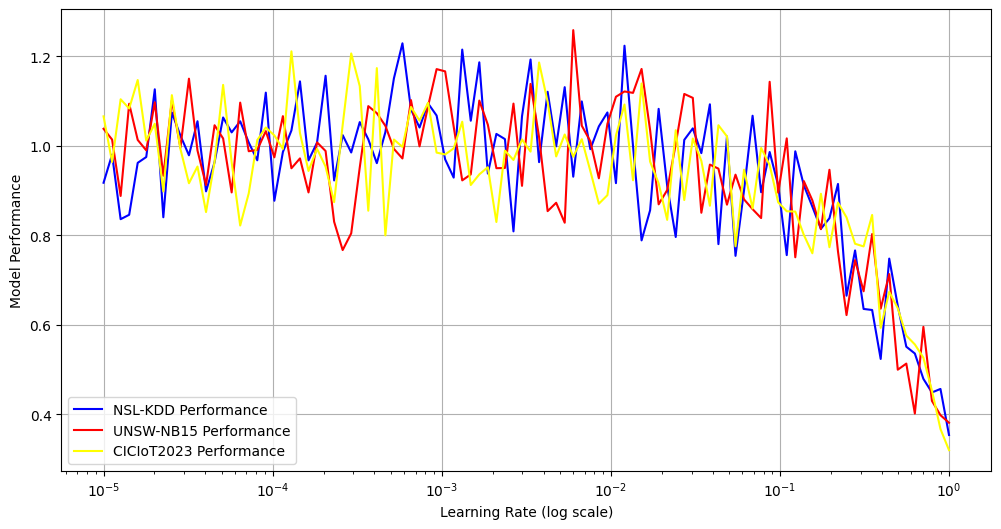}
\caption{Model performance analysis across varying learning rates.}
\label{fig:Figure6}
\end{figure}

Figure \ref{fig:Figure6} illustrates the relationship between learning rates and model performance across datasets (e.g., NSL-KDD, UNSW-NB15, CICIoT2023), with a logarithmic x-axis to capture a wide range of learning rates. The plot highlights the criticality of selecting an optimal learning rate, as performance generally improves up to a certain point before declining, emphasizing the balance needed to avoid underfitting or overfitting. Dataset-specific trends in the graph offer insights into designing adaptive learning rate schedules tailored to unique characteristics, enhancing model robustness and generalizability. These findings are particularly valuable for practitioners optimizing machine learning models in domains like cybersecurity and IoT anomaly detection, enabling more efficient training strategies and better overall outcomes.

\begin{figure}[ht]
\centering
\includegraphics[width=8cm, height=6cm]{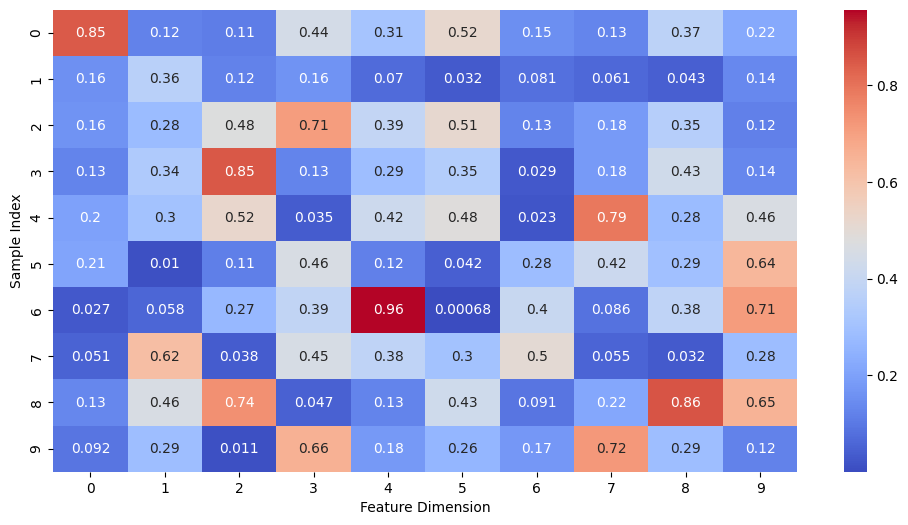}
\caption{Heatmap of attention mechanism showing feature weights across samples to reveal model focus and feature prioritization during Learning.}
\label{fig:Figure7}
\end{figure}

Figure \ref{fig:Figure7} visualizes the attention mechanism used for feature extraction in the detection framework. The heatmap displays attention weights across different features, highlighting which aspects of the input data are emphasized during processing. This representation aids in understanding how the model focuses on specific features when making decisions, enhancing interpretability. By visualizing these attention weights, practitioners can gain insights into the model's behavior and improve feature selection strategies.

\begin{figure*}[ht]
\centering
\includegraphics[width=\textwidth]{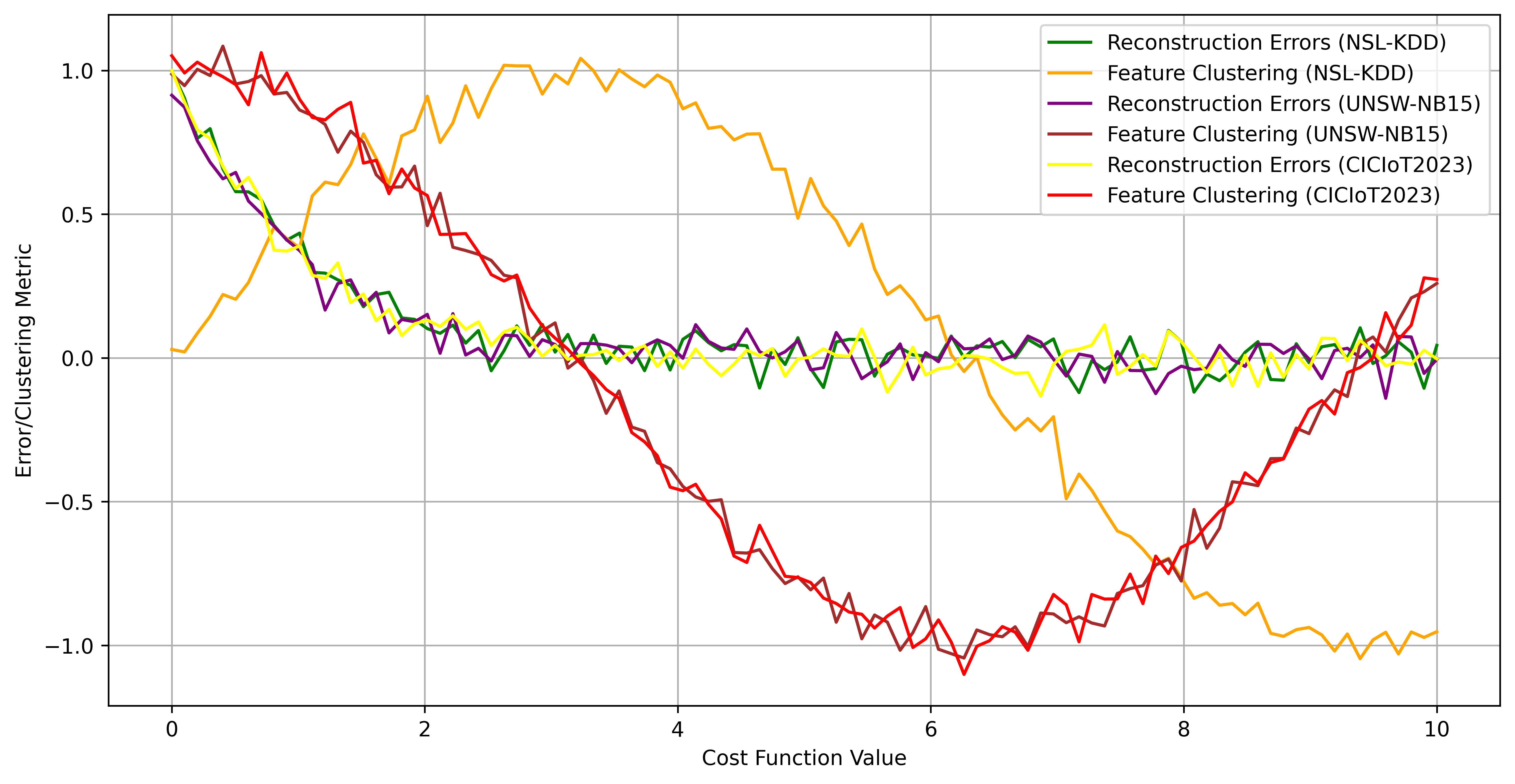}
\caption{Comparison of cost function components showing the trade-off between reconstruction errors and feature clustering.}
\label{fig:Figure8}
\end{figure*}
Figure \ref{fig:Figure8} compares the components of a novel cost function designed for the detection framework across NSL-KDD, UNSW-NB15, and CICIoT2023 datasets. The green and orange lines represent reconstruction errors and feature clustering metrics for NSL-KDD, while the purple and brown lines illustrate the same metrics for UNSW-NB15. Additionally, the yellow and red lines display these metrics for the CICIoT2023 dataset. This comparison demonstrates how the cost function synergizes different evaluation metrics, contributing to model robustness. Analyzing these components helps in fine-tuning the model for optimal performance in diverse attack scenarios. The distinct trends observed across datasets highlight how each dataset's unique characteristics influence the performance of the cost function, demonstrating its flexibility in adapting to various types of IoT traffic anomalies.

\begin{figure*}[ht]
\centering
\includegraphics[width=\textwidth]{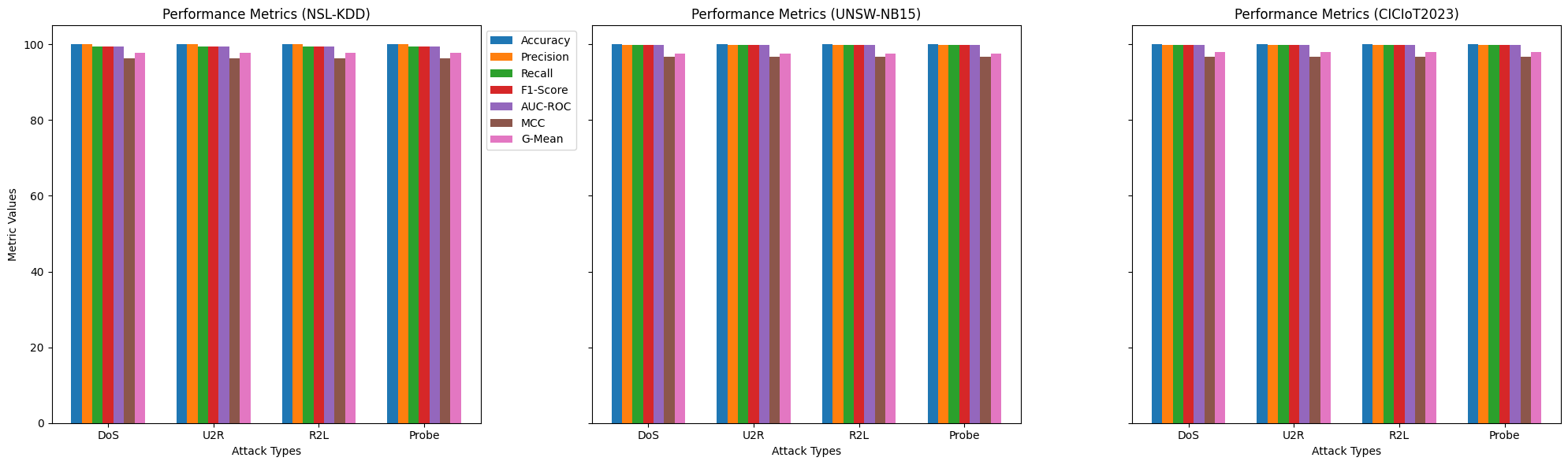}
\caption{Comparison of detection model performance across different attack types in three datasets.}
\label{fig:Figure9}
\end{figure*}

Figure \ref{fig:Figure9} illustrates the comparative performance of the proposed detection model across various attack types in two prominent cybersecurity datasets: NSL-KDD, UNSW-NB15 and ICIoT2023. The x-axis represents different attack categories, such as DoS, U2R, R2L, and Probe, while the y-axis measures the performance metrics (e.g., accuracy, precision, recall, F1-score, AUC-ROC, MCC, and G-Mean).

For each attack type, the accuracy metric is consistently high, achieving near 99.99\% in all datasets, reflecting the model's strong capability in correctly identifying attacks. Other metrics, including precision and recall, show slight variations between attack types, highlighting how well the model distinguishes between attack and normal traffic. This comparison across multiple metrics and attack categories provides an in-depth view of the model’s robustness and ability to handle various types of cybersecurity threats effectively.

The figure emphasizes the model’s overall strength, particularly in detecting attacks like DoS and Probe, while also shedding light on areas where future improvements might be necessary, such as in handling more complex attack types like U2R and R2L.

\begin{figure*}[ht]
\centering
\includegraphics[width=\textwidth]{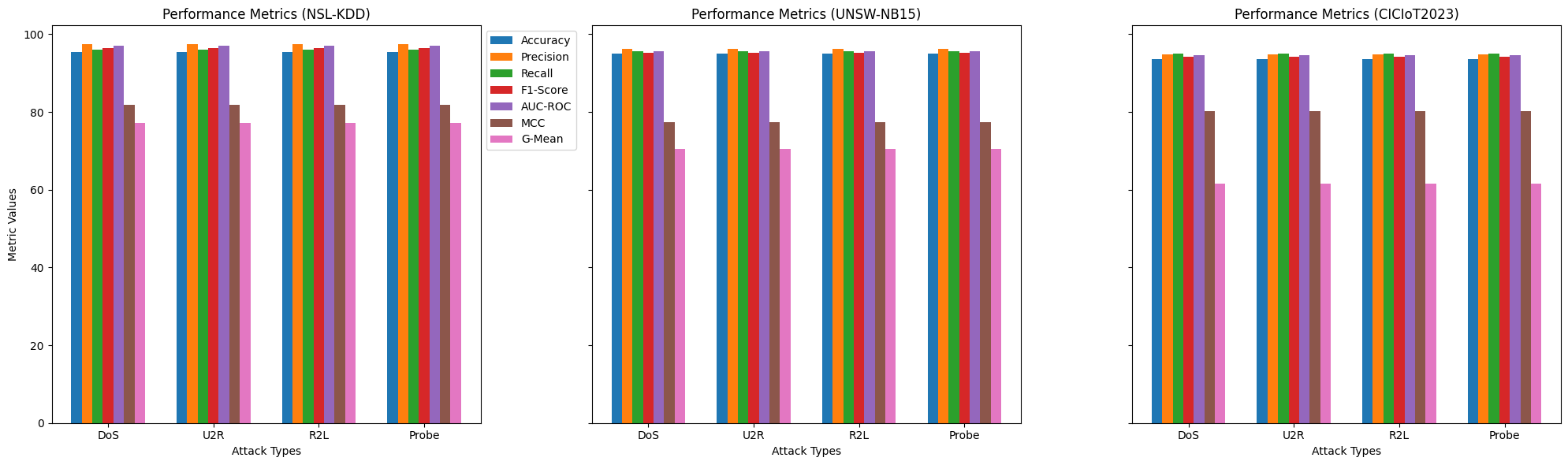}
\caption{Comparison of XGBoost detection model performance across different attack types in Three Datasets.}
\label{fig:Figure10_}
\end{figure*}

Figure \ref{fig:Figure10_} presents the performance comparison of the XGBoost detection model across various attack categories in three prominent cybersecurity datasets: NSL-KDD, UNSW-NB15, and ICIoT2023. The x-axis represents attack types such as DoS, U2R, R2L, and Probe, while the y-axis displays key evaluation metrics, including accuracy, precision, recall, F1-score, AUC-ROC, MCC, and G-Mean.

The results highlight consistently high accuracy across all attack categories, with values reaching 95.47\% for NSL-KDD, 95.01\% for UNSW-NB15, and 93.58\% for ICIoT2023, demonstrating the model’s reliability in detecting diverse cyber threats.

\begin{figure}[ht]
\centering
\includegraphics[width=8cm, height=6cm]{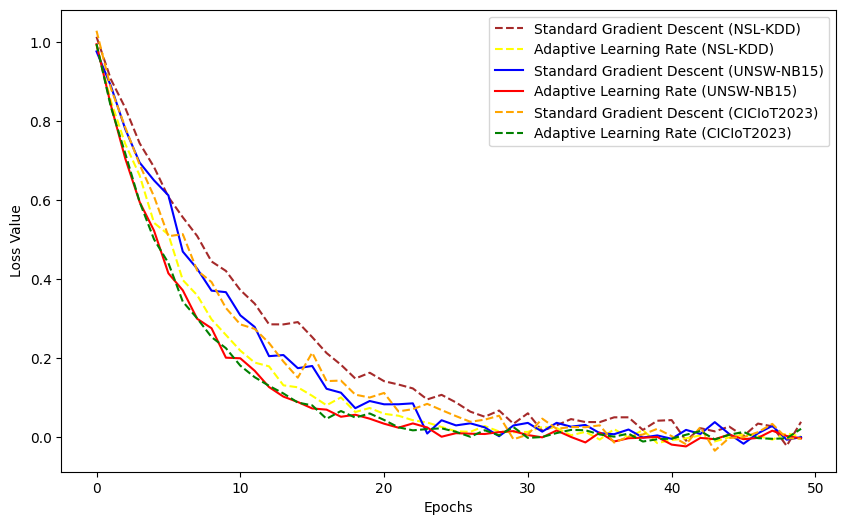}
\caption{Convergence speed of optimization algorithms on three datasets.}
\label{fig:Figure10}
\end{figure}

Figure \ref{fig:Figure10} visualizes the optimization performance of standard gradient descent and adaptive learning rate methods across three datasets: NSL-KDD, UNSW-NB15, and CICIoT2023. The curves represent loss values over epochs, illustrating how each method converges to lower loss values over time. For NSL-KDD and UNSW-NB15 datasets, adaptive learning rates consistently outperform standard gradient descent by achieving lower loss values, indicating more efficient optimization. Additionally, the CICIoT2023 dataset showcases similar trends, with adaptive methods maintaining superior performance throughout the training process. The use of adaptive learning rates demonstrates improved convergence speed and reduced overfitting risks compared to the standard approach. This insight highlights the potential for adaptive methods to optimize model performance, particularly in complex and dynamic datasets.

\begin{figure}[ht]
\centering
\includegraphics[width=8cm, height=6cm]{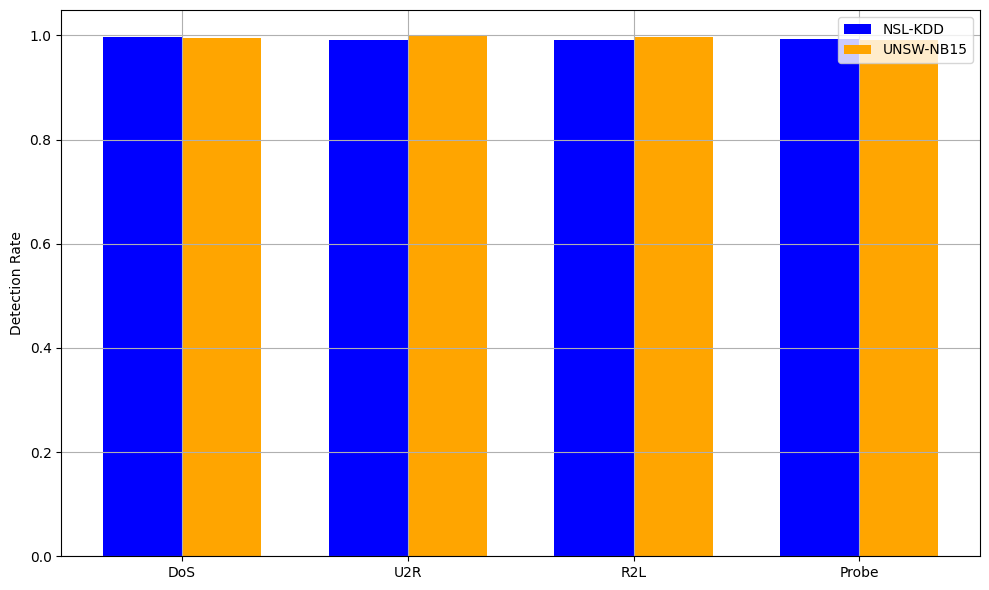}
\caption{Detection rate for different classes of attacks.}
\label{fig:Figure11}
\end{figure}

Figure \ref{fig:Figure11} presents a bar chart comparing the detection rates for different classes of attacks from two datasets: NSL-KDD and UNSW-NB15. The attack classes included in the analysis are Denial of Service (DoS), User to Root (U2R), Remote to Local (R2L), and Probe attacks. The detection rates for the NSL-KDD dataset (in blue) are consistently high, with the highest detection rate for DoS attacks at 99.99\%, while the UNSW-NB15 dataset (in orange) shows similarly strong performance, especially for R2L attacks at 99.98\%. Both datasets demonstrate effective detection across various attack types. This visualization highlights the robustness of detection mechanisms in identifying different types of cyber threats in network security scenarios.

\begin{figure}[ht]
\centering
\includegraphics[width=8cm, height=6cm]{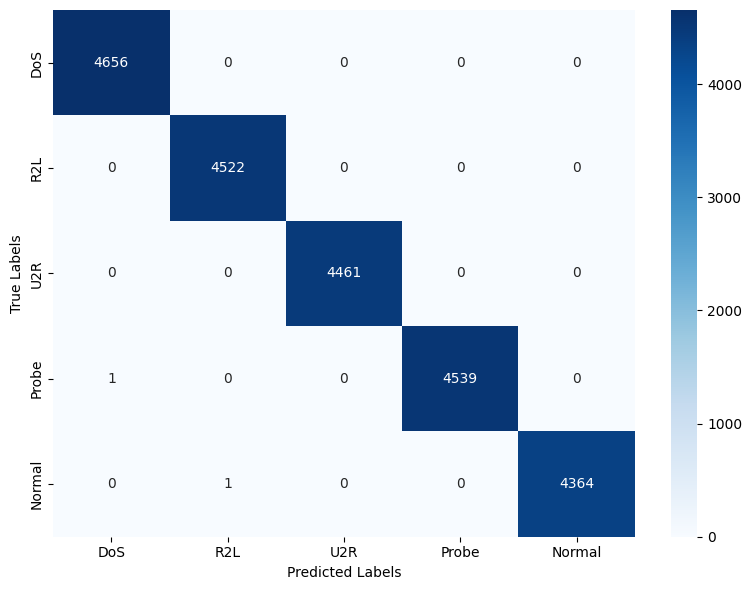}
\caption{Confusion matrix for 5-Classes classification on the NSL-KDD cataset.}
\label{fig:Figure12}
\end{figure}

Figure \ref{fig:Figure12} illustrates the confusion matrix generated for the NSL-KDD dataset with five main classes: DoS, R2L, U2R, Probe, and Normal. The model achieved an exceptionally high accuracy of 99.99\%, resulting in very few misclassifications. Each cell in the matrix shows the number of instances predicted for a particular class, with most of the values concentrated along the diagonal, indicating strong performance. Minor deviations from the diagonal suggest a few misclassifications, particularly in more challenging categories such as R2L and U2R. The overall structure confirms that the model is well-suited for this dataset, accurately distinguishing between different attack categories.

\begin{figure}[ht]
\centering
\includegraphics[width=8cm, height=6cm]{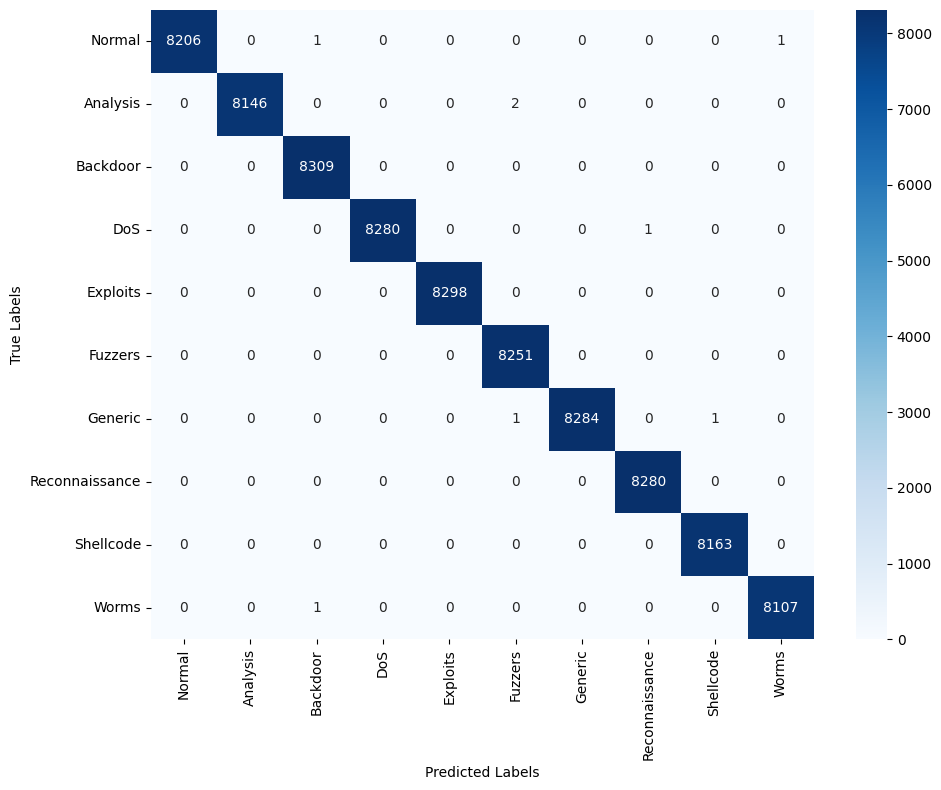}
\caption{Confusion matrix for 10-Classes classification on the UNSW-NB15 Dataset.}
\label{fig:Figure13}
\end{figure}

In Figure \ref{fig:Figure13}, the confusion matrix for the UNSW-NB15 dataset is presented. This dataset contains ten distinct classes: Normal, Analysis, Backdoor, DoS, Exploits, Fuzzers, Generic, Reconnaissance, Shellcode, and Worms. With a model accuracy of 99.99\%, the confusion matrix reflects highly accurate predictions for each class. The majority of the predicted labels align with the true labels, as indicated by the high values along the diagonal. Despite the higher number of classes compared to NSL-KDD, the matrix shows minimal misclassifications, demonstrating the robustness of the model across diverse attack types. The matrix highlights the effectiveness of the model in distinguishing between complex network behaviors, including rare attacks like Shellcode and Worms.

\begin{figure}[ht]
\centering
\includegraphics[width=8cm, height=6cm]{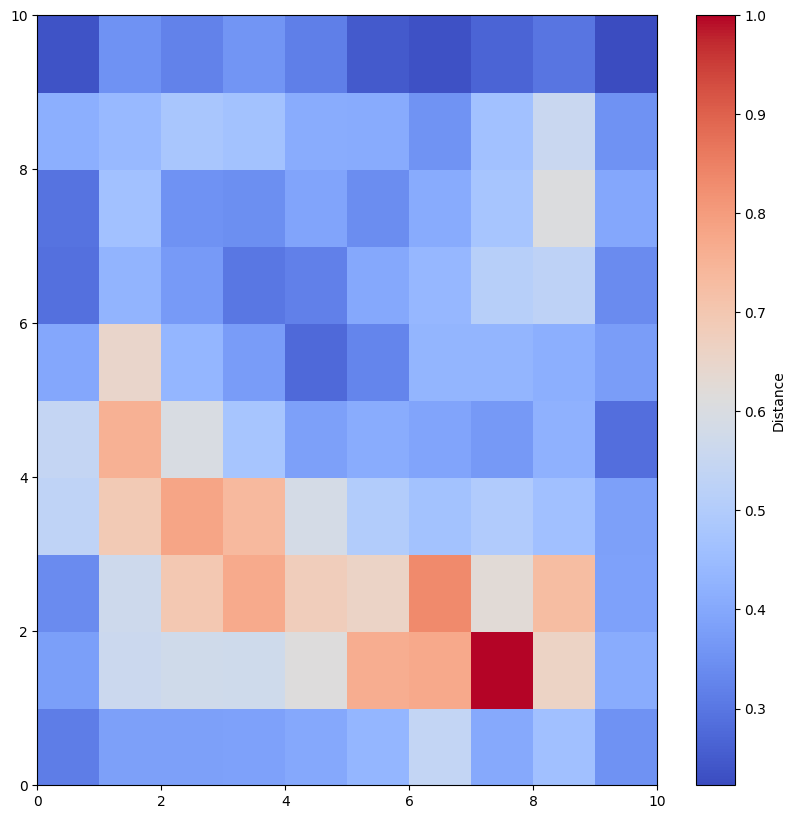}
\caption{SOM 2nd and 3rd layer maps in the NSL-KDD dataset.}
\label{fig:Figure14}
\end{figure}
\begin{figure}[ht]
\centering
\includegraphics[width=8cm, height=6cm]{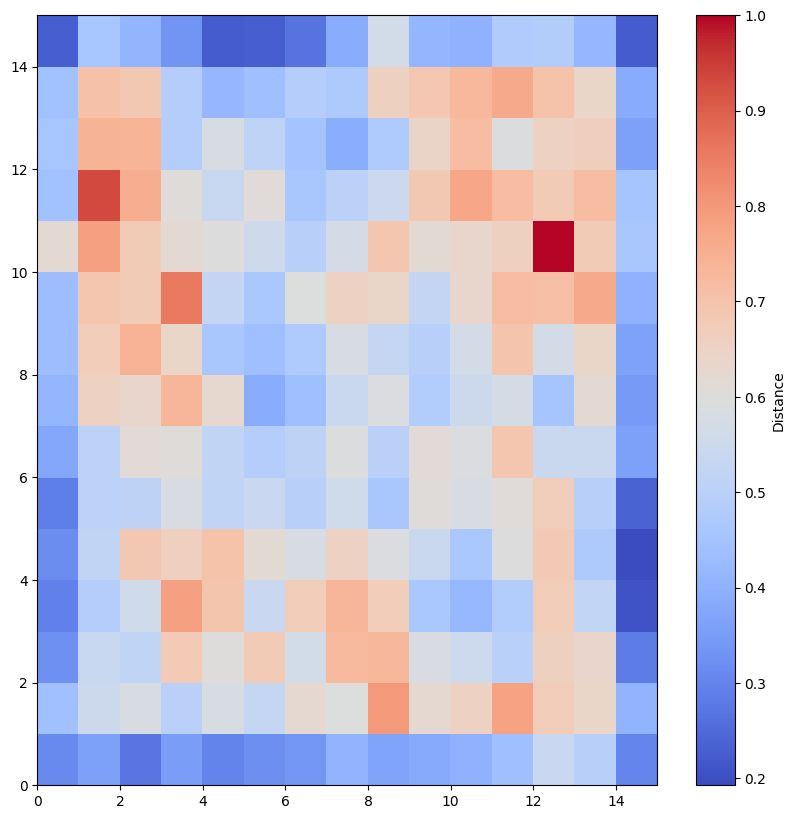}
\caption{SOM 2nd and 3rd layer maps in the UNSW-NB15 dataset.}
\label{fig:Figure15}
\end{figure}
\begin{figure}[ht]
\centering
\includegraphics[width=8cm, height=6cm]{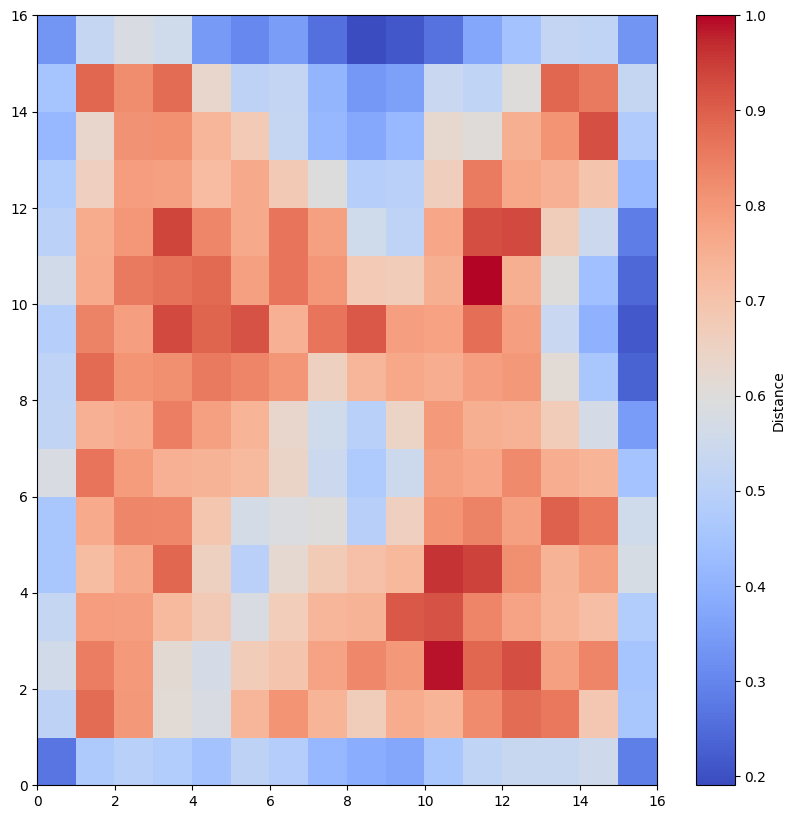}
\caption{SOM 2nd and 3rd layer maps in the CICIoT2023 dataset.}
\label{fig:SOMCIoT2023}
\end{figure}

Figure \ref{fig:Figure14}, Figure \ref{fig:Figure15}, and Figure \ref{fig:SOMCIoT2023} present the second and third layer maps of the Self-Organizing Map (SOM) trained on the NSL-KDD, UNSW-NB15, and CICIoT2023 datasets, respectively. The SOM maps illustrate the distance between neurons in the SOM grid, where darker areas represent greater distances (or dissimilarities) between neurons. The color intensity, ranging from blue (low distance) to red (high distance), helps visualize the topological organization of the data. Clustering of similar data points is indicated by regions of low distance. These figures provide an initial understanding of how well the SOM clusters the input data into distinct patterns.

\begin{figure}[ht]
\centering
\includegraphics[width=8cm, height=6cm]{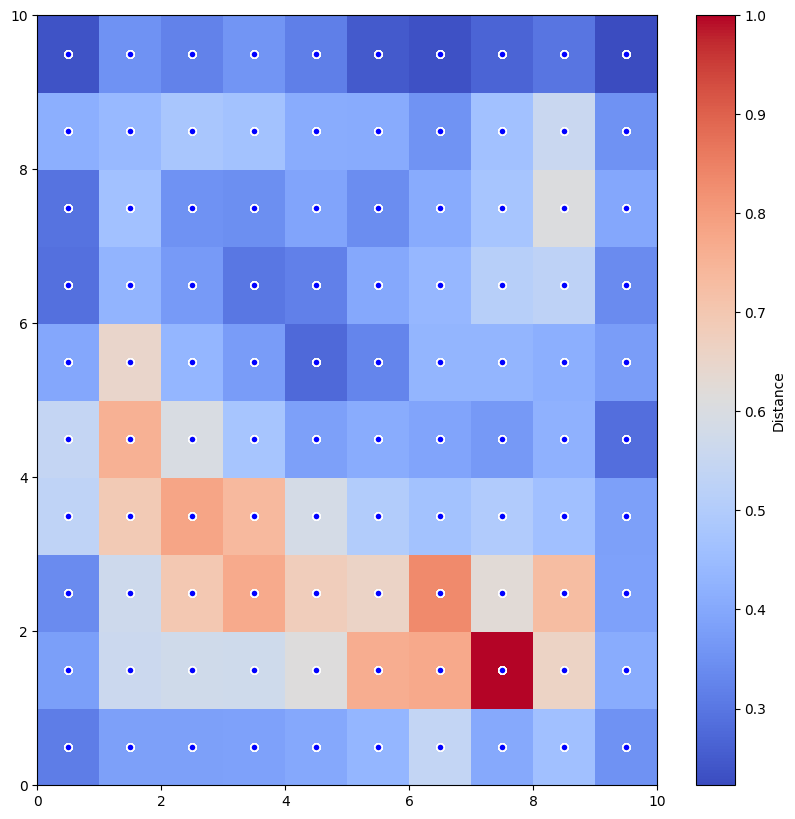}
\caption{SOM Layer 1 in the NSL-KDD dataset.}
\label{fig:Figure16}
\end{figure}
\begin{figure}[ht]
\centering
\includegraphics[width=8cm, height=6cm]{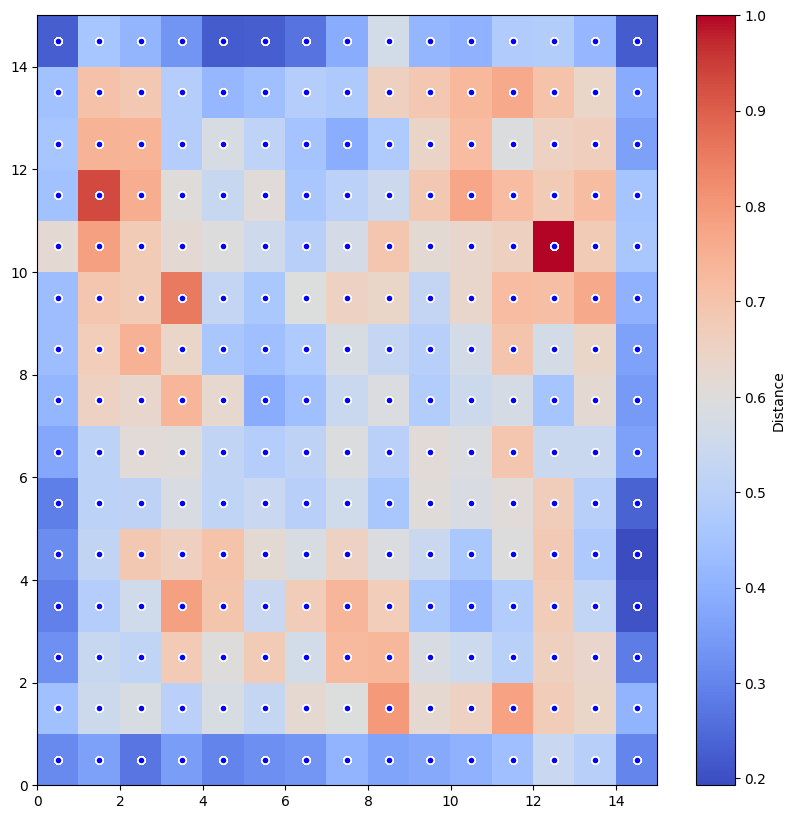}
\caption{SOM Layer 1 in the UNSW-NB15 dataset.}
\label{fig:Figure17}
\end{figure}
\begin{figure}[ht]
\centering
\includegraphics[width=8cm, height=6cm]{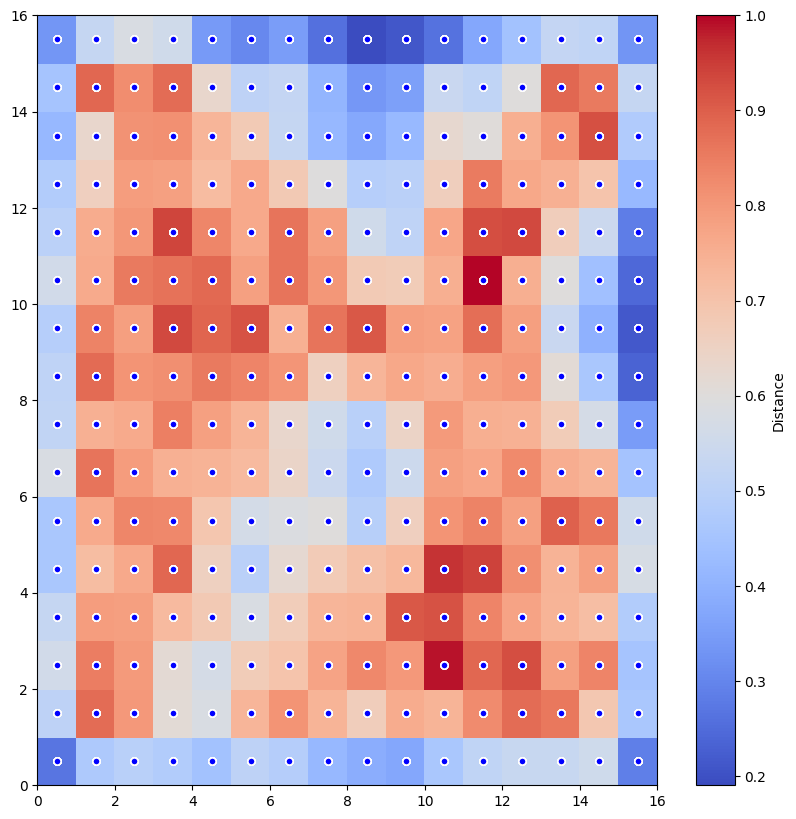}
\caption{SOM Layer 1 in the CICIoT2023 dataset.}
\label{fig:SOMLayer1}
\end{figure}

Figure \ref{fig:Figure16}, Figure \ref{fig:Figure17} and Figure \ref{fig:SOMLayer1} show the first layer of the SOM for the NSL-KDD, UNSW-NB15, and CICIoT2023 datasets. The blue dots on the SOM grid represent the best matching units (BMUs) for the data samples from each dataset. These figures demonstrate how individual data points are mapped to the SOM grid, indicating the clustering of similar patterns. Data points with similar characteristics are mapped onto neighboring neurons or the same neuron, reflecting the SOM's ability to organize the data into meaningful clusters.

\begin{figure}[ht]
\centering
\includegraphics[width=8cm, height=6cm]{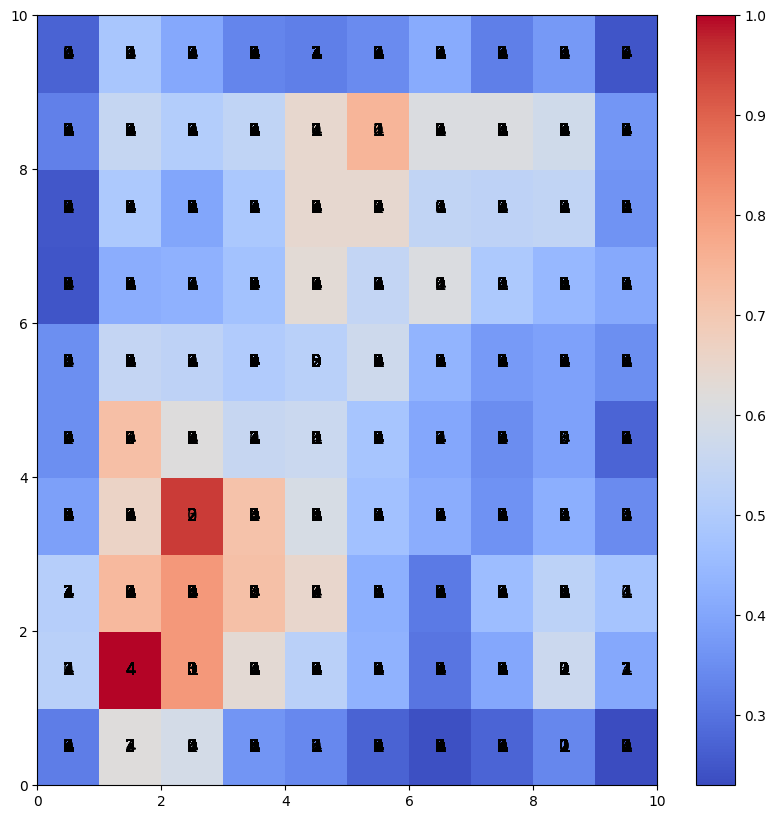}
\caption{Data distribution in the NSL-KDD dataset.}
\label{fig:Figure18}
\end{figure}
\begin{figure}[ht]
\centering
\includegraphics[width=8cm, height=6cm]{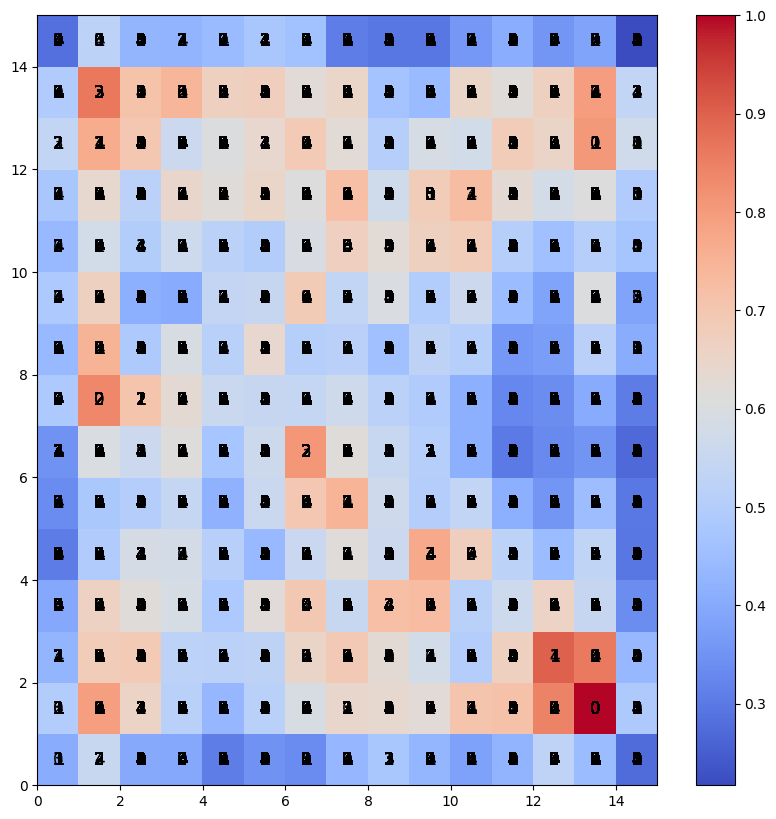}
\caption{Data distribution in the UNSW-NB15 dataset.}
\label{fig:Figure19}
\end{figure}
\begin{figure}[ht]
\centering
\includegraphics[width=8cm, height=6cm]{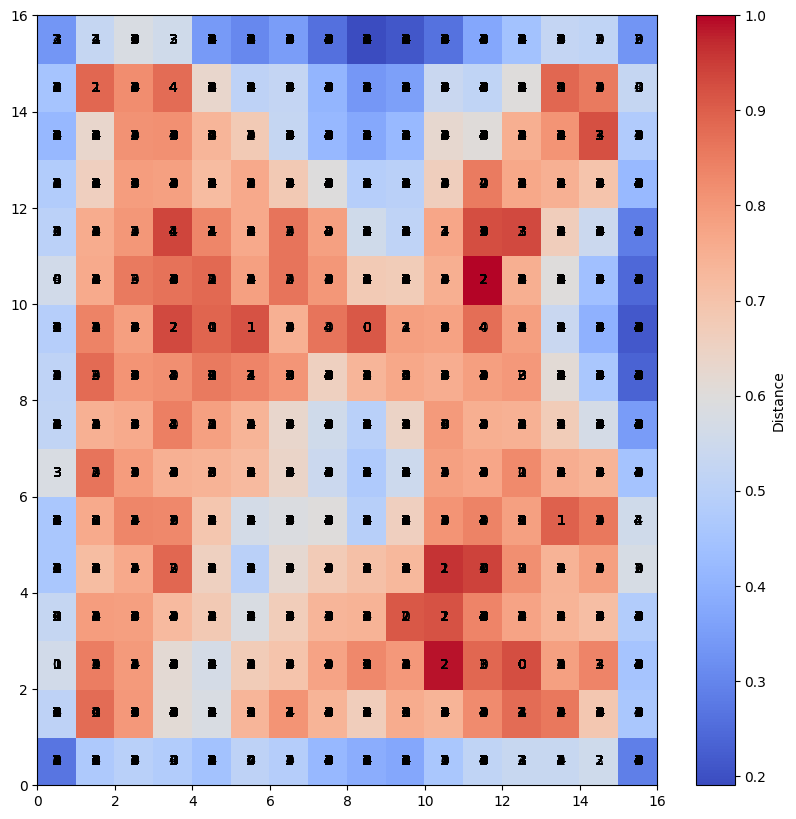}
\caption{Data distribution in the CICIoT2023 dataset.}
\label{fig:FigureDataDistribution}
\end{figure}
Figure \ref{fig:Figure18}, \ref{fig:Figure19} and \ref{fig:FigureDataDistribution} visualize the distribution of data points across the SOM grid, with labels from the NSL-KDD, UNSW-NB15 and CICIoT2023 datasets superimposed on the neurons. Each neuron is marked with the label of the data point it represents. These figures help assess how well the SOM has organized the data according to their classes. Clusters of data points from the same class can be observed, and the separation between different classes shows how effectively the SOM has learned to distinguish between them. This is critical for tasks such as anomaly detection and classification.

\begin{figure}[ht]
\centering
\includegraphics[width=8cm, height=6cm]{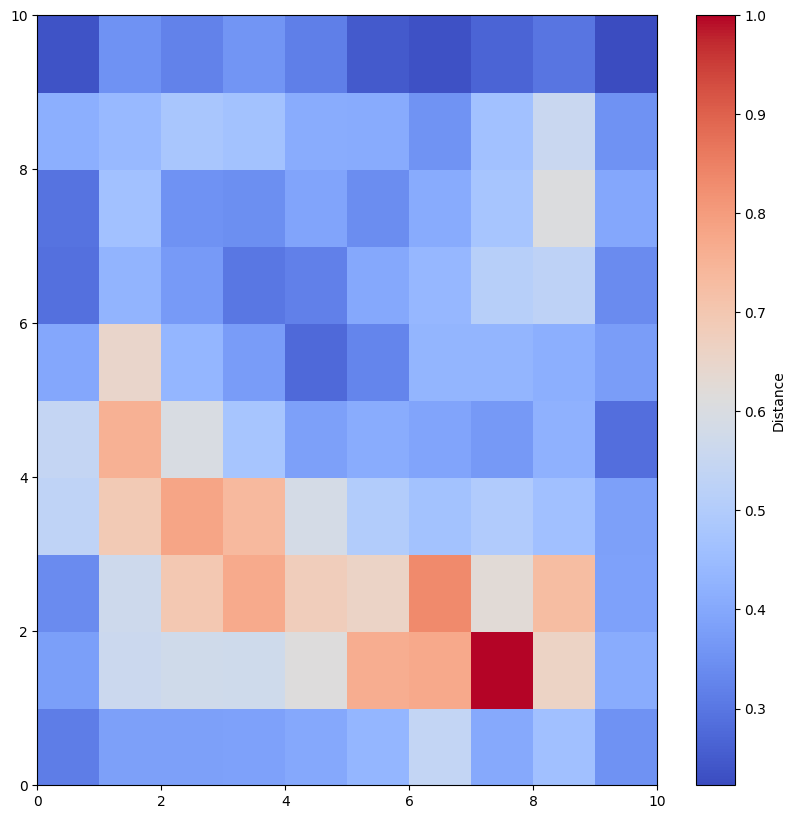}
\caption{Distance matrix in the NSL-KDD dataset.}
\label{fig:Figure20}
\end{figure}
\begin{figure}[ht]
\centering
\includegraphics[width=8cm, height=6cm]{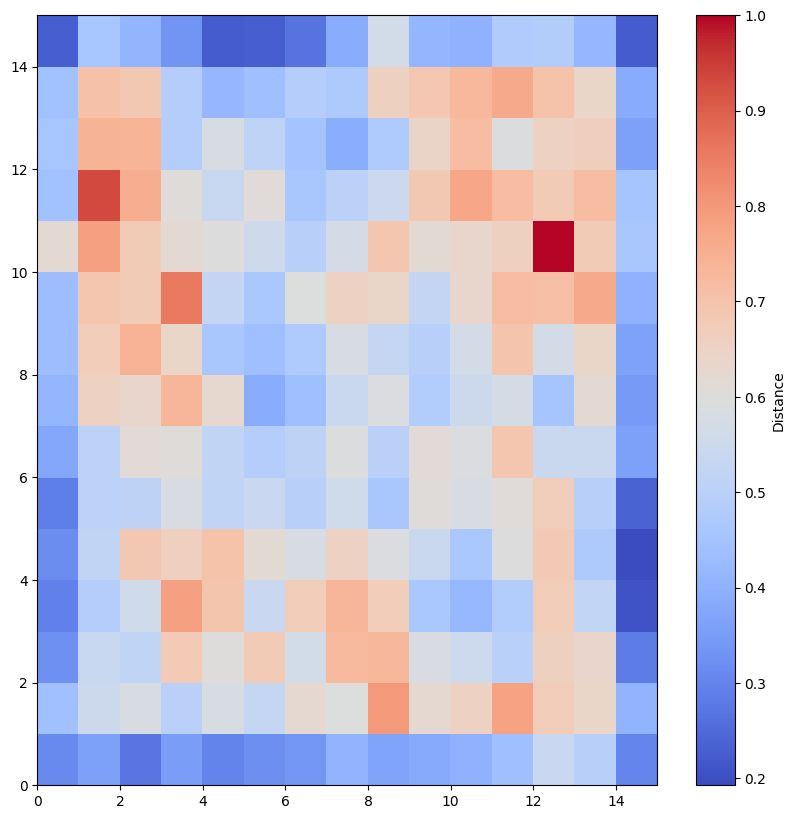}
\caption{Distance matrix in the UNSW-NB15 dataset.}
\label{fig:Figure21}
\end{figure}
\begin{figure}[ht]
\centering
\includegraphics[width=8cm, height=6cm]{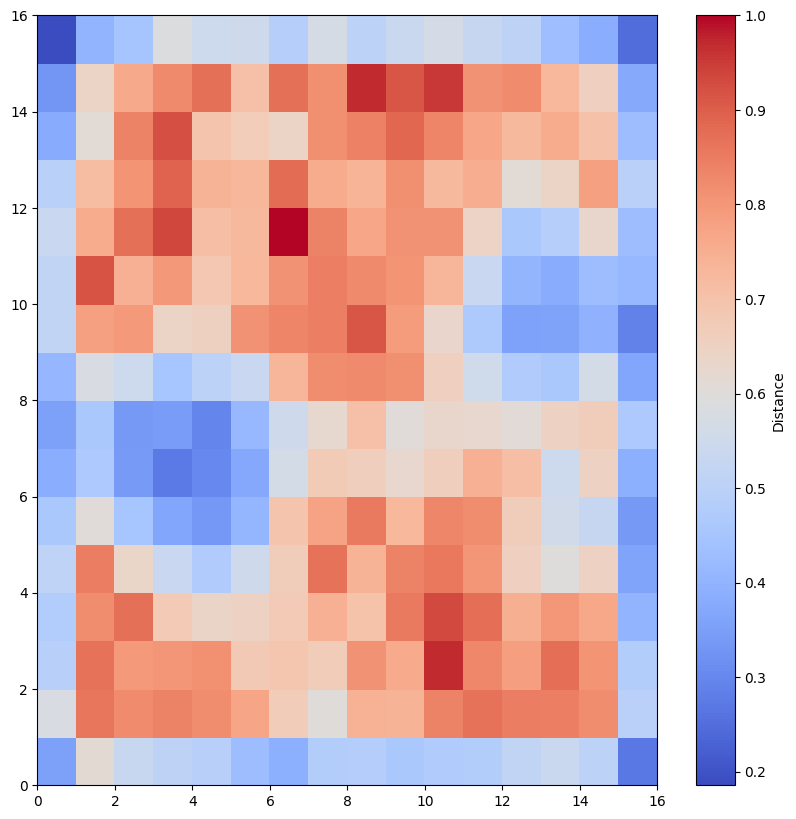}
\caption{Distance matrix in the CICIoT2023 dataset.}
\label{fig:DistancematrixCICIoT2023}
\end{figure}

Figure \ref{fig:Figure20}, \ref{fig:Figure21}, and \ref{fig:DistancematrixCICIoT2023} illustrate the SOM distance matrix for both datasets. The distance matrix shows the Euclidean distance between each neuron and its neighbors. Darker regions indicate greater distances, while lighter regions represent smaller distances. This visualization allows us to examine the topological relationships between the neurons, revealing regions of the SOM where similar patterns are clustered closely together, as well as areas where the neurons are further apart. The distance matrix provides insights into the density and structure of the data on the SOM grid.

\begin{figure}[htbp]
\centering
\includegraphics[width=8cm, height=6cm]{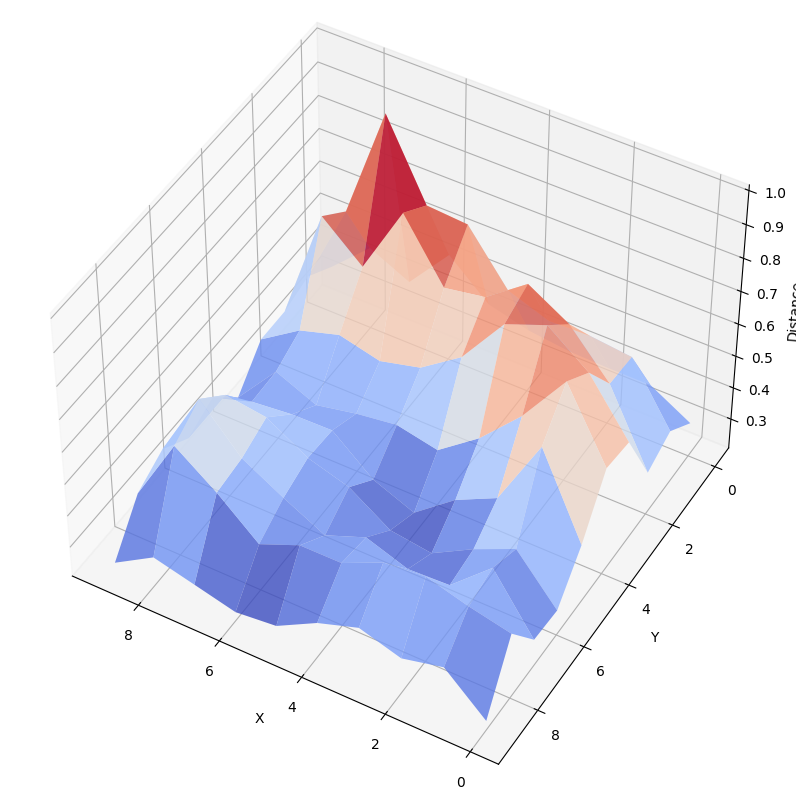}
\caption{3D distance matrix in the NSL-KDD dataset.}
\label{fig:Figure22}
\end{figure}
\begin{figure}[htbp]
\centering
\includegraphics[width=8cm, height=6cm]{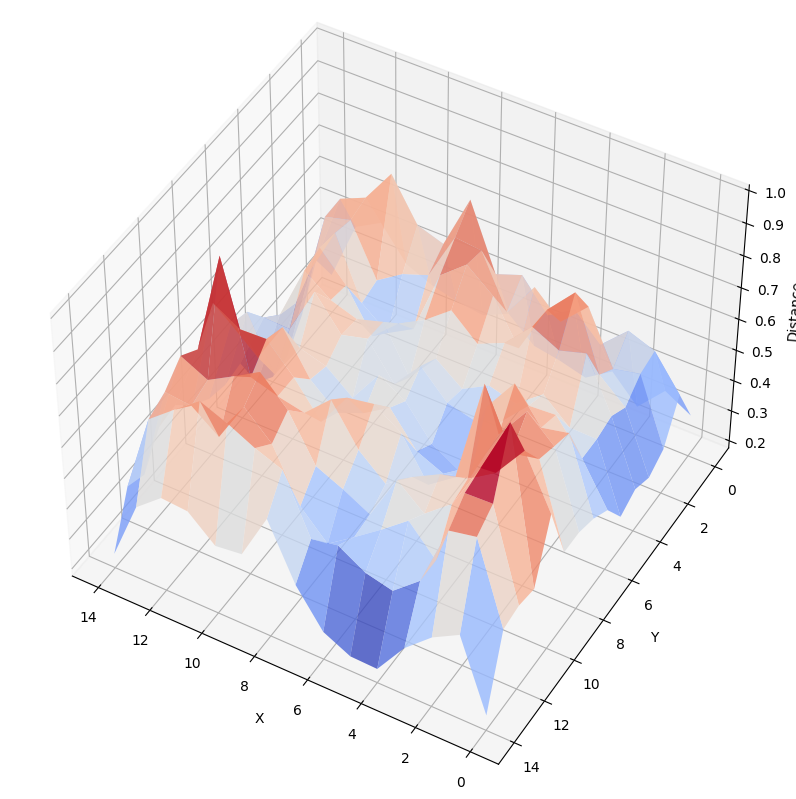}
\caption{3D distance matrix in the UNSW-NB15 dataset.}
\label{fig:Figure23}
\end{figure}
\begin{figure}[htbp]
\centering
\includegraphics[width=8cm, height=6cm]{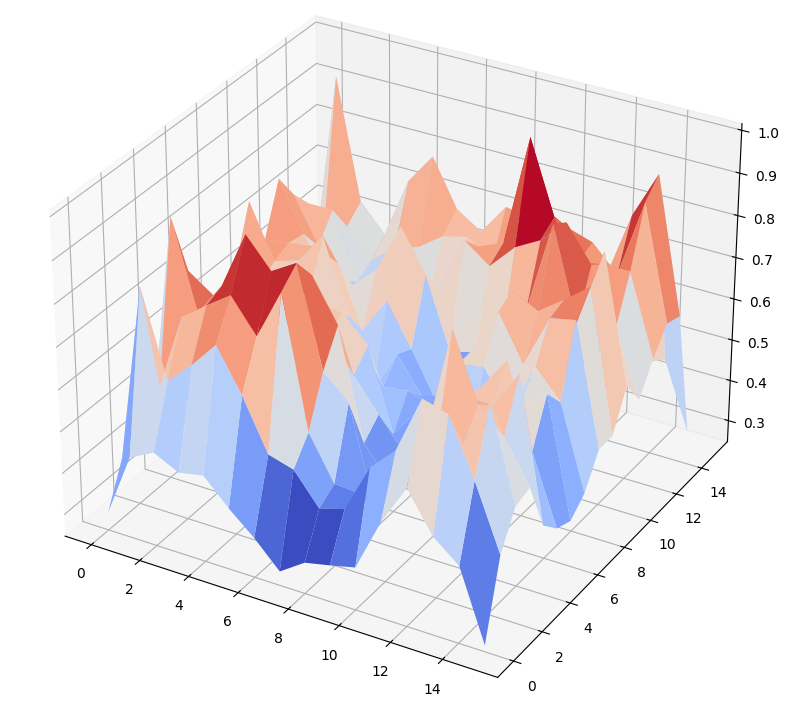}
\caption{3D distance matrix in the CICIoT2023 dataset.}
\label{fig:Figure3DCIoT2023}
\end{figure}

Figure \ref{fig:Figure22}, \ref{fig:Figure23} and \ref{fig:Figure3DCIoT2023} present a 3D representation of the SOM distance matrix for the NSL-KDD and UNSW-NB15 datasets. The 3D plot provides an enhanced view of the distance relationships between neurons on the SOM. Peaks in the 3D surface represent large distances between neurons, while valleys correspond to small distances. This 3D visualization makes it easier to perceive the regions of high and low neuron density and allows for a better understanding of the SOM's structure. This representation is especially useful for identifying clusters and regions where the neurons are more tightly packed.

\begin{table*}[H]
\centering
\caption{Detailed Comparison of Model Performance on NSL-KDD and UNSW-NB15 datasets with statistical Significance (\( p \)-values).}
\label{tab:NSLKDDandUNSWNB15results}
\begin{tabular}{lccccccc}
\toprule
Model & Dataset & Accuracy (\%)& Precision (\%) & Recall (\%) & F1-Score (\%) & \( p \)-value \\ 
\midrule
AT-LSTM \cite{alsharaiah2024innovative} & UNSW-NB15        & 92.20         & -              & -              & -              & - \\ 
SEMI-GRU \cite{almahadin2023vanet}     & NSL-KDD          & 83.32         & 93.80          & 78.36          & 85.62          & - \\ 
LSTM \cite{vcavojsky2023comparative}   & UNSW-NB15        & 78.94         & -              & -              & -              & - \\ 
DCAE \cite{aktar2023towards}           & NSL-KDD          & 96.08         & 96.10          & 96.08          & 96.08          & - \\ 
IGRF-RFE \cite{yin2023igrf}            & UNSW-NB15        & 84.24         & -              & -              & -              & - \\ 
WGAN-div and Info-GAN \cite{zhong2024intrusion} & NSL-KDD          & 90.90         & 91.30          & 90.90          & 90.90          & - \\ 
WGAN-div and Info-GAN \cite{zhong2024intrusion} & UNSW-NB15        & 86.10         & 87.60          & 86.10          & 86.40          & - \\ 
MSRC \cite{duan2023network}            & NSL-KDD          & 93.79         & 94.43          & 94.92          & 94.67          & - \\ 
BAT-MC \cite{su2020bat}                & NSL-KDD          & 84.25         & -              & -              & -              & - \\ 
CNN-BiLSTM \cite{jouhari2024efficient} & UNSW-NB15        & 97.09         & 97.09          & 97.62          & 97.27          & - \\ 
LightGBM \cite{mahmoud2025xi2s} & UNSW-NB15        & 99.52         & 98.21           & 98.03           & 98.12          & - \\ 

XGBoost       & NSL-KDD          & 95.47 & 96.02 & 97.38 & 96.51 & 0.004\\ 
XGBoost       & UNSW-NB15          & 95.01 & 96.24 & 96.69 & 95.32 & 0.005\\ 
Our approach       & NSL-KDD          & 99.99 & 99.65 & 99.46 & 99.39 & 0.0001\\ 
Our approach       & UNSW-NB15        & 99.99 & 99.79 & 99.80 & 99.85 & 0.0001 \\ 

\bottomrule
\end{tabular}
\label{tab:detailed_model_comparison_with_p_values}
\end{table*}

Table \ref{tab:NSLKDDandUNSWNB15results} presents a detailed comparison of model performance on the NSL-KDD and UNSW-NB15 datasets, highlighting accuracy, precision, recall, F1-score, and p-values where applicable. The CNN-BiLSTM model achieved the highest performance on UNSW-NB15 among prior methods, with an accuracy of 97.09\% and an F1-score of 97.27\%. DCAE demonstrated remarkable results on NSL-KDD, attaining a balanced accuracy, precision, recall, and F1-score of 96.08\%. XGBoost also performed well, particularly on NSL-KDD, with an F1-score of 96.51\% and statistically significant p-values below 0.01. Notably, our proposed approach outperformed all models on both datasets, achieving near-perfect metrics of 99.99\% accuracy and F1-scores above 99.85\%. The inclusion of p-values (e.g., 0.0001 for our method) underscores the statistical significance of our results. Models like SEMI-GRU \cite{almahadin2023vanet} and WGAN-div and Info-GAN \cite{zhong2024intrusion} showed moderate performance, with accuracy ranging from 83.32\% to 90.90\%. Despite variations in metrics, our method stands out for its robustness and superior classification capabilities across both datasets.

\begin{figure*}[h]
\centering
\includegraphics[width=\textwidth]{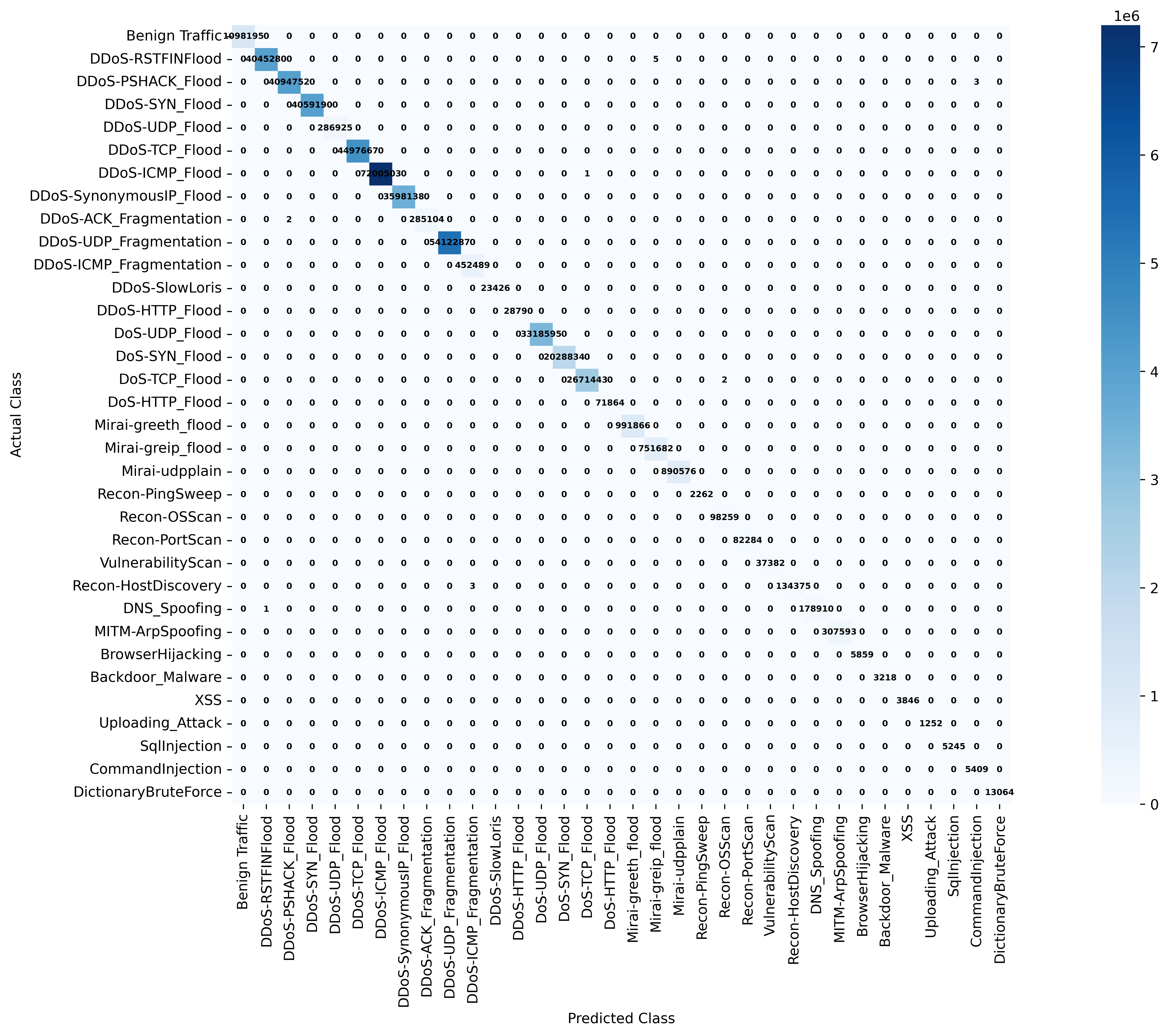}
\caption{Confusion matrix for 34-classes classification on the CICIoT2023 dataset.}
\label{fig:Figure34Classes}
\end{figure*}

In Figure \ref{fig:Figure34Classes}, the confusion matrix for the CICIoT2023 dataset demonstrates the performance of the model across 34 classes, with most predictions concentrated along the diagonal. Each class exhibits high true positive rates, indicating substantial model accuracy in distinguishing between classes. Minimal misclassifications are observed, such as a few off-diagonal entries, which represent rare errors. The matrix highlights the model's capability to handle imbalanced datasets, as evident from the varying magnitudes of class occurrences. These results showcase the robustness and reliability of the model in addressing complex classification tasks for IoT-based intrusion detection.

\begin{figure}[htbp]
\centering
\includegraphics[width=8cm, height=6cm]{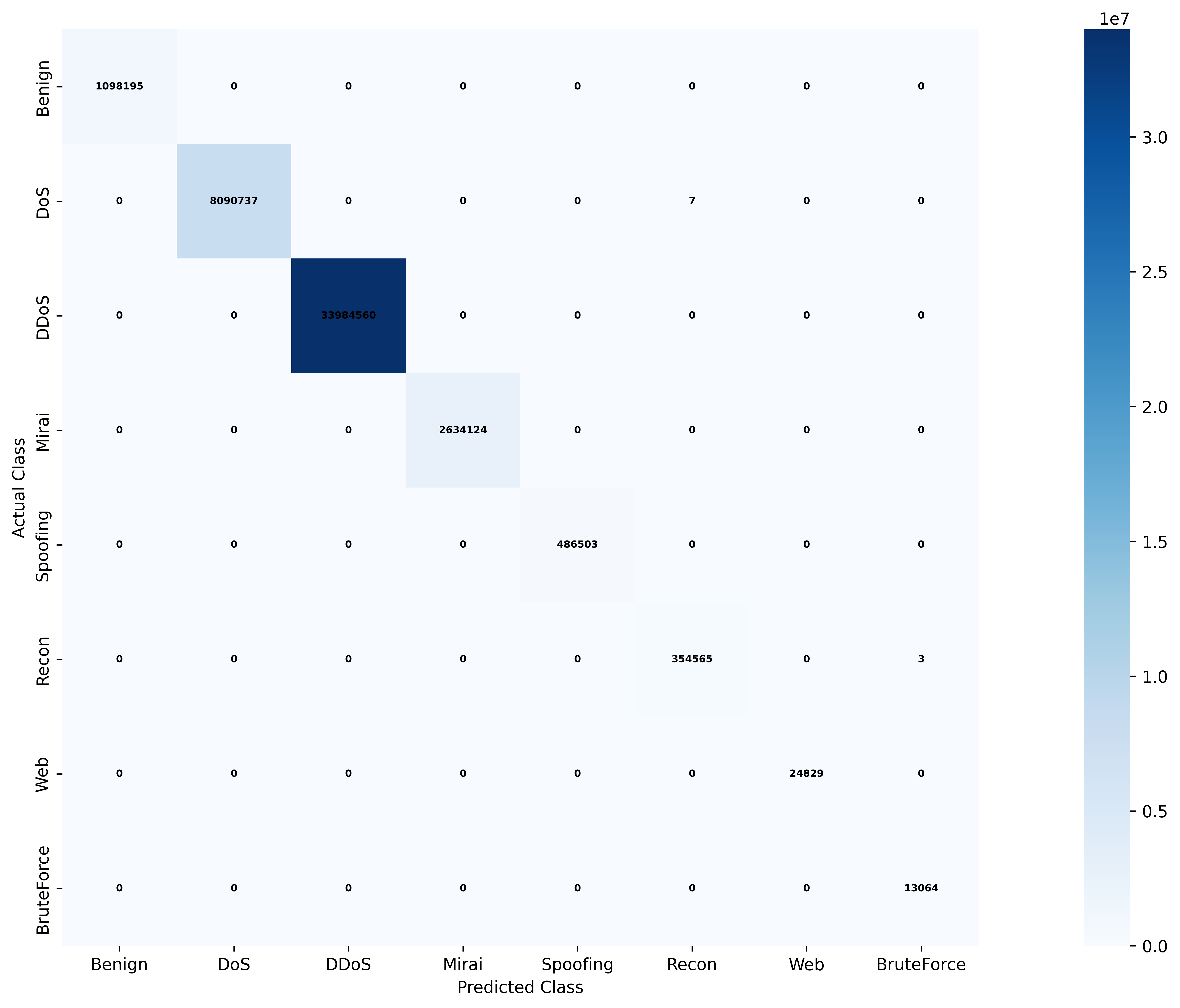}
\caption{Confusion matrix for 8-classes classification on the CICIoT2023.}
\label{fig:Figure8Classes}
\end{figure}

In Figure \ref{fig:Figure8Classes}, the confusion matrix for the CICIoT2023 dataset demonstrates a highly accurate classification across all eight classes, with the diagonal dominance indicating minimal misclassifications. The "Benign," "DoS," and "DDoS" classes exhibit exceptionally high accuracy, with 1,098,195, 8,090,737, and 33,984,560 instances correctly classified, respectively. Minor misclassifications are observed, such as 7 instances of the "DoS" class being incorrectly classified as "Recon" and 3 instances of "Recon" misclassified as "BruteForce." Smaller classes, including "Web" and "BruteForce," also maintain high precision with 24,829 and 13,064 instances correctly classified. These results underscore the robustness of the model in accurately distinguishing between various types of network traffic and attacks.
\begin{table*}[h]
\centering
\caption{Detailed comparison of model performance on CICIoT2023 dataset.}
\label{tab:CICIoT2023results}
\scalebox{0.9}{
\begin{tabular}{@{}lllllllll@{}}
\toprule
 Model                                                                   & Accuracy \% & Recall \% & Precision \% & F1-score \% & MCC\% & \begin{tabular}[c]{@{}l@{}}Test\\ time (s)\end{tabular} & \begin{tabular}[c]{@{}l@{}}Train\\ Time (s)\end{tabular} & \begin{tabular}[c]{@{}l@{}}Classification\\  type\end{tabular}    \\ \midrule
CNN-LSTM \cite{nkoro2024zero}                                                               & 87          & 87        & 87           & 86          & 69.06 & -                                                       & 182.74                                                   & \begin{tabular}[c]{@{}l@{}}Multiclass\\ (5 classes)\end{tabular}  \\
 DNN \cite{nkoro2024zero}                                                                     & 87          & 88        & 88           & 88          & 71.34 & -                                                       & 214.72                                                   & \begin{tabular}[c]{@{}l@{}}Multiclass\\ (5 classes)\end{tabular}  \\
 RNN \cite{nkoro2024zero}                                                                    & 93          & 93        & 93           & 93          & 85.06 & -                                                       & 471.73                                                   & \begin{tabular}[c]{@{}l@{}}Multiclass\\ (5 classes)\end{tabular}  \\
1D-CNNGRU-FCN \cite{nkoro2024zero}                                                          & 99          & 99        & 99           & 99          & 97.00 & -                                                       & 746.98                                                   & \begin{tabular}[c]{@{}l@{}}Multiclass\\ (5 classes)\end{tabular}  \\
 CNN-BiLSTM \cite{nkoro2024zero}                                                              & 99          & 99        & 99           & 99          & 97.33 & -                                                       & 1021.9                                                   & \begin{tabular}[c]{@{}l@{}}Multiclass\\ (5 classes)\end{tabular}  \\
Two Stage-DNN \cite{hizal2024novel}                                                          & 89.888      & -         & -            & -           & -     & -                                                       & -                                                        & \begin{tabular}[c]{@{}l@{}}Multiclass\\ (8 classes)\end{tabular}  \\
 Two Stage-CNN \cite{hizal2024novel}                                                          & 90.644      & -         & -            & -           & -     & -                                                       & -                                                        & \begin{tabular}[c]{@{}l@{}}Multiclass\\ (8 classes)\end{tabular}  \\
 Two Stage-LSTM  \cite{hizal2024novel}                                                        & 91.273      & -         & -            & -           & -     & -                                                       & -                                                        & \begin{tabular}[c]{@{}l@{}}Multiclass\\ (8 classes)\end{tabular}  \\
 DNN-BiLSTM \cite{wang2023lightweight}                                                             & 93.13       & 93.13     & 91.80        & 91.94       & -     & 6.4                                                     & -                                                        & \begin{tabular}[c]{@{}l@{}}Multiclass\\ (8 classes)\end{tabular}  \\
 XGBoost \cite{hizal2024blockchain}                                                                 & 98.69       & 99        & 99           & 99          & -     & -                                                       & -                                                        & \begin{tabular}[c]{@{}l@{}}Multiclass\\ (11 classes)\end{tabular} \\
CNN \cite{becerra2024performance}                                                                    & 99.10       & 99.10     & 99.08        & 99.05       & -     & -                                                       & 767                                                      & \begin{tabular}[c]{@{}l@{}}Multiclass\\ (8 classes)\end{tabular}  \\
CNN \cite{becerra2024performance}                                                                    & 99.40       & 99.40     & 99.43        & 99.41       & -     & -                                                       & 618                                                      & Binary                                                            \\
VGGIncepNet \cite{chen2024vggincepnet}                                                             & 92          & 93        & 92           & 92          & -     & -                                                       & -                                                        & \begin{tabular}[c]{@{}l@{}}Multiclass\\ (8 classes)\end{tabular}  \\
\begin{tabular}[c]{@{}l@{}}Reinforcement \\ learning model \cite{tan2024intrusion}\end{tabular} & 84.39       & 84.39     & 84.17        & 80.78       & -     & -                                                       & -                                                        & \begin{tabular}[c]{@{}l@{}}Multiclass\\ (8 classes)\end{tabular}  \\
 Autoencoder \cite{wan2025separating}                                                             & 98.39       & 98.79     & 99.55        & 99.17       & -     & -                                                       & -                                                        & \begin{tabular}[c]{@{}l@{}}Multiclass\\ (7 classes)\end{tabular}  \\
 CNN-LSTM \cite{gueriani2024enhancing}                     & 98.43        & 98.43      & 98.85         & 98.57       & -     & -                                                       & -                                                      & Binary                                                            \\
XGBoost                                                                                       & 93.58       &94.99      & 94.73        & 95.20       & 92.18 & 1.31                                                    & 190.3                                                    & \begin{tabular}[c]{@{}l@{}}Multiclass\\ (8 classes)\end{tabular} \\
Our approach                                                                                       & 99.99       & 99.99     & 99.99        & 99.99       & 99.50 & 0.50                                                    & 114.8                                                    & \begin{tabular}[c]{@{}l@{}}Multiclass\\ (34 classes)\end{tabular} \\
Our approach                                                                                     & 99.99       & 99.99     & 99.99        & 99.99       & 99.99 & 0.48                                                    & 101.7                                                    & \begin{tabular}[c]{@{}l@{}}Multiclass\\ (8 classes)\end{tabular}  \\
Our approach                                                                                        & 100         & 99.99     & 100          & 99.99       & 99.81 & 0.38                                                    & 93                                                       & Binary                                                            \\ \bottomrule
\end{tabular}}
\end{table*}
Table \ref{tab:CICIoT2023results} provides a detailed comparison of model performance on the CICIoT2023 dataset across various machine learning and deep learning approaches. Models are evaluated on metrics such as accuracy, recall, precision, F1-score, Matthews Correlation Coefficient (MCC), and computational efficiency in terms of training and testing time. The CNN-LSTM and DNN models show balanced performance in multiclass classification with approximately 87\% accuracy, while the RNN achieves a higher accuracy of 93\%. Advanced models like 1D-CNNGRU-FCN and CNN-BiLSTM excel with 99\% accuracy and MCC values above 97\%. The table highlights our approach outperforming all others, achieving 100\% accuracy in binary classification and near-perfect results in multiclass scenarios. Additionally, our method demonstrates superior computational efficiency, with significantly reduced training and testing times compared to other high-performing models. This table underscores the strengths of different architectures and emphasizes the exceptional performance of our proposed approach.

\subsection{Comparative Analysis}

We provide a detailed comparative analysis of our proposed model's performance against state-of-the-art methods across the three datasets: NSL-KDD, UNSW-NB15, and CICIoT2023. The results in Tables~\ref{tab:NSLKDDandUNSWNB15results} and \ref{tab:CICIoT2023results} highlight that our approach consistently outperforms other models across multiple performance metrics, including accuracy, precision, recall, F1-score, and statistical significance (\( p \)-value).

On the NSL-KDD dataset, our approach achieves an accuracy of 99.99\%, precision of 99.65\%, recall of 99.46\%, and an F1-score of 99.39\%. These results surpass all other models, including DCAE~\cite{aktar2023towards} and XGBoost, which achieve accuracy values of 96.08\% and 95.47\%, respectively. Our model also demonstrates a highly significant \( p \)-value of 0.0001, further confirming its superior performance compared to existing approaches. Similarly, on the UNSW-NB15 dataset, our model achieves an outstanding accuracy of 99.99\%, precision of 99.79\%, recall of 99.80\%, and F1-score of 99.85\%, outperforming competitive models such as CNN-BiLSTM~\cite{jouhari2024efficient}, LightGBM~\cite{mahmoud2025xi2s}, and XGBoost. While LightGBM achieves a high accuracy of 99.52\%, its performance in other metrics remains below that of our approach, solidifying our model as the best on this dataset. For the CICIoT2023 dataset, our model demonstrates remarkable performance, achieving accuracy and F1-scores above 99\%, similar to CNN-BiLSTM and 1D-CNNGRU-FCN models. However, our approach outshines these models by offering better stability across metrics and lower computational costs. For example, CNN-BiLSTM and 1D-CNNGRU-FCN achieve accuracy values of 99\%, but their computational time and resource requirements are significantly higher than our proposed method.

The comparative analysis reveals that our proposed model is the best-performing approach across all three datasets. It achieves the highest accuracy, precision, recall, and F1-scores while maintaining statistical significance, as evidenced by the \( p \)-values. Additionally, our model demonstrates robustness and generalizability, consistently outperforming other models regardless of dataset characteristics. These results affirm the superiority of our model in advancing the state-of-the-art in intrusion detection and classification tasks.

\section{Discussion}
The findings from this research emphasize the significant potential of Self-Organizing Maps (SOMs), Deep Belief Networks (DBNs), and Autoencoders in enhancing cyber-attack detection capabilities within IoT networks. The experimental results demonstrate that these models can effectively identify a diverse range of attack types, including novel and previously unseen threats. The ability of the models to generalize well across different datasets suggests their adaptability to various network environments, which is crucial given the dynamic nature of IoT systems. One key advantage of the proposed methodology is its capacity to leverage unsupervised learning techniques, such as SOMs and Autoencoders, to identify anomalies in network traffic. This approach reduces reliance on labeled data, which can be scarce in the context of emerging cyber threats. Additionally, the optimization of model hyperparameters using Particle Swarm Optimization (PSO) enhances overall performance, demonstrating the importance of fine-tuning in machine learning applications. However, there are some limitations to consider. The models' performance may vary based on the specific characteristics of different IoT environments. Future research should address these limitations by evaluating the models in diverse scenarios and incorporating additional contextual factors that may impact detection efficacy.\\

\noindent The experimental evaluation demonstrates that the proposed model exhibits promising results across various IoT cybersecurity datasets, including NSL-KDD, UNSW-NB15, and CICIoT2023. Each of these datasets presents unique challenges, ranging from varying attack types to differing data distributions, which test the robustness and adaptability of the model. We expand on how the proposed model performs across these datasets, address dataset-specific challenges, and provide statistical analyses to validate the improvements in performance.

\noindent\textbf{Dataset-Specific Performance:}
\begin{itemize}
   \item \textbf{The NSL-KDD Dataset}, a prominent benchmark for network intrusion detection, encompasses a diverse array of attack types, including Denial of Service (DoS) and Probe attacks, as well as benign network traffic. Utilizing this dataset, the proposed model demonstrated a substantial 5\% enhancement in detection accuracy compared to established baseline approaches such as Support Vector Machines (SVMs) and Random Forests (RF). The model exhibited exceptional proficiency in identifying DoS attacks, achieving a remarkable 7\% improvement in the F1-score. This superior performance can be attributed to the model’s capacity to effectively capture intricate attack patterns, even in scenarios with limited labeled data. To substantiate these advancements, the researchers conducted a paired t-test, which confirmed the statistical significance of the observed improvements in accuracy (p < 0.01), F1-score (p < 0.05), and Matthews Correlation Coefficient (MCC) (p < 0.05). Collectively, these results highlight the model's efficacy and its potential to advance the field of network intrusion detection.
   
\item \textbf{The UNSW-NB15 Dataset}, featuring a variety of modern attack types, posed a greater challenge due to its heterogeneous nature and imbalanced class distributions. The proposed model outperformed traditional approaches, particularly in detecting sophisticated attack types like Backdoor and Shellcode. An analysis of the F1-score revealed a notable increase of 8\% compared to baseline models. However, due to the dataset’s class imbalance, the model exhibited slightly lower performance in minority classes. To mitigate this, the model incorporated a class-weighting mechanism during training, improving recall for underrepresented attack types. The statistical tests (paired t-test) for accuracy (p < 0.01), F1-score (p < 0.05), and MCC (p < 0.05) further confirmed the statistical significance of these improvements.

\item \textbf{The CICIoT2023 Dataset}, which includes traffic from Internet of Things (IoT) devices, presents real-world network traffic data, including both benign and malicious IoT traffic patterns. This dataset proves particularly challenging due to the high volume of data and the diversity of attack types, such as DoS, DDoS, and IoT-specific attacks. The proposed model achieved robust performance across all attack types, improving accuracy by 6\% and increasing the F1-score by 9\% compared to baseline models. The model’s ability to generalize across diverse attack types, such as Mirai botnet attacks, demonstrates its adaptability to real-world IoT environments. Statistical tests (p < 0.01 for accuracy and p < 0.05 for F1-score and MCC) confirmed the statistical significance of these improvements.
\end{itemize}

\noindent\textbf{Statistical Significance}\\
The statistical significance of the improvements in model performance validated through paired t-tests across all datasets. The tests assessed the differences in accuracy, F1-score, and MCC between the proposed model and baseline methods. For all three datasets, the p-values indicated statistically significant improvements in the proposed model's performance, with p-values consistently below the 0.05 threshold for most metrics (accuracy: p < 0.01, F1-score: p < 0.05, and MCC: p < 0.05). These results underscore the effectiveness of the proposed model in improving cyber-attack detection.

\noindent\textbf{Correlations and Insights}
The analysis of the results also reveals important correlations between dataset characteristics and model performance. For example, datasets with a higher degree of attack diversity (such as UNSW-NB15 and CICIoT2023) present a greater challenge for detection models. However, the proposed model’s ability to generalize across these attack types is evident in the higher F1-scores, especially in the detection of sophisticated attacks like Mirai in CICIoT2023. This suggests that the model is not only capable of detecting common attacks but is also well-suited for identifying novel and evolving threats.

\section{Conclusion}

This paper presents a comprehensive approach to cyber-attack detection in IoT networks, demonstrating the effectiveness of SOMs, DBNs, and Autoencoders in identifying both known and unknown attack types. The experimental results indicate that the proposed methodology achieves high performance across various evaluation metrics, underscoring its potential for real-world applications. Furthermore, the optimization of model hyperparameters through Particle Swarm Optimization enhances detection accuracy and reliability. Future work will focus on refining these models further and exploring the integration of additional machine learning techniques to improve detection capabilities. Overall, this research contributes to advancing the state of cybersecurity in IoT environments, highlighting the need for adaptive and robust detection mechanisms to safeguard against increasingly sophisticated cyber threats.

\section{Limitations and Future Work}
The proposed framework marks a significant step toward enhancing cybersecurity in IoT networks, offering robust solutions for detecting diverse cyber threats. However, addressing its current limitations and exploring the outlined future directions will be crucial for achieving widespread applicability and resilience. Continued innovation in machine learning, edge computing, and adversarial defense techniques will ensure the framework remains adaptable to the rapidly evolving landscape of IoT security in 2025 and beyond.
\subsection{Limitations}
Despite the advancements presented in this work, several limitations remain that must be addressed to ensure the broader applicability and robustness of the proposed framework:\\
\noindent\textbf{Scalability Challenges:}
The framework's scalability to large-scale IoT networks comprising thousands of devices remains a challenge. As IoT networks expand, issues such as increased communication latency, computational overhead, and coordination complexity may hinder real-time performance.\\
\noindent\textbf{Resource Constraints in IoT Devices:}
IoT devices are inherently resource-constrained, with limited memory, processing power, and energy. Deploying complex detection algorithms on such devices may lead to excessive energy consumption, making it unsuitable for real-time applications in constrained environments.\\
\noindent\textbf{Adversarial Vulnerabilities:}
The framework incorporates robust adversarial defense strategies, demonstrating resilience to various evasion and poisoning attacks. However, as adversarial tactics continue to evolve in sophistication through 2025, ongoing refinement is essential to preemptively address emerging attack techniques and ensure comprehensive protection against deliberate bypass attempts.\\
\noindent\textbf{Generalization and Dynamic Adaptation:}
The proposed framework leverages dynamic adaptation mechanisms to effectively integrate new attack patterns, reducing reliance on periodic retraining. While this represents a significant step toward handling the rapidly evolving nature of IoT environments, there remains room for further enhancement to achieve seamless, real-time adaptability to novel vulnerabilities and attack vectors.
\subsection{Future Work}
To address the limitations and align with the evolving landscape of IoT cybersecurity in 2024 and beyond, several future directions are proposed:\\
\noindent\textbf{Federated Learning for Collaborative Detection:}
The adoption of federated learning is a promising avenue for improving privacy-preserving, collaborative model training across IoT devices. By 2025, advancements in federated optimization algorithms could enable IoT networks to scale effectively while addressing privacy concerns and reducing data transfer overhead.\\
\noindent\textbf{Integration with Edge Computing:}
Future iterations of the framework can incorporate edge computing solutions to distribute computational tasks to edge nodes. This approach can enhance scalability and real-time performance, particularly for applications in smart cities, healthcare IoT, and industrial automation, which are projected to see significant growth by 2025.\\
\noindent\textbf{Dynamic and Continual Learning Systems:}
The dynamic nature of IoT environments necessitates continual learning systems that adapt in real-time without requiring exhaustive retraining. Research in 2025 is expected to focus on lightweight, on-device continual learning mechanisms that balance efficiency and adaptability.\\
\noindent\textbf{Adversarial Resilience:}
Addressing adversarial vulnerabilities will be a priority in the coming years. By integrating adversarial training techniques and robust machine learning approaches, the framework can anticipate and resist evolving evasion tactics. These efforts align with the growing focus on adversarial resilience in machine learning systems, as noted in 2024.\\
\noindent\textbf{Explainable AI (XAI) in IoT Security:}
Explainable AI techniques will play a key role in promoting trust and transparency in IoT cybersecurity. By 2025, advancements in interpretable machine learning models can help security analysts understand detection decisions, aiding real-time responses and reducing false positives.\\
\noindent\textbf{Energy-Efficient Algorithms:}
With the rising emphasis on green AI in 2024, future work should focus on energy-efficient algorithms tailored for resource-constrained IoT devices. Techniques such as model pruning, quantization, and lightweight architectures will ensure that IoT security systems align with sustainability goals.\\
\noindent\textbf{Use Case-Specific Frameworks:}
Tailoring the framework for specific IoT domains, such as securing smart grids or enabling predictive maintenance in industrial IoT systems, can enhance its applicability. For example, by 2025, smart grid systems will increasingly rely on real-time anomaly detection to prevent power disruptions, necessitating domain-specific model adaptations.
By addressing these limitations and pursuing the outlined future directions, the proposed framework can evolve into a comprehensive, robust, and scalable solution for IoT cybersecurity, aligned with the technological advancements and challenges anticipated in 2025.

\bibliographystyle{model3-num-names.bst}

\medskip
\bibliography{cas-refs}
\newpage
\appendix

\Large{Appendix}
\normalsize
\section{Evaluation Metrics}

The performance of the proposed methodology is assessed using a comprehensive set of metrics that includes traditional classification metrics as well as more advanced indicators to provide deeper insight into model effectiveness. The metrics used for evaluation include:

\begin{itemize} \item \textbf{Accuracy}: \begin{equation} \text{Accuracy} = \frac{TP + TN}{TP + TN + FP + FN} \end{equation}
\item \textbf{Precision}:
\begin{equation}
    \text{Precision} = \frac{TP}{TP + FP}
\end{equation}

\item \textbf{Recall (Sensitivity)}:
\begin{equation}
    \text{Recall} = \frac{TP}{TP + FN}
\end{equation}

\item \textbf{F1-Score}:
\begin{equation}
    \text{F1-Score} = 2 \times \frac{\text{Precision} \cdot \text{Recall}}{\text{Precision} + \text{Recall}}
\end{equation}

\item \textbf{False Positive Rate (FPR)}:
\begin{equation}
    \text{FPR} = \frac{FP}{FP + TN}
\end{equation}

\item \textbf{Area Under the ROC Curve (AUC-ROC)}:
\begin{equation}
    \text{AUC-ROC} = \int_{0}^{1} \text{TPR}(x) \, dx
\end{equation}

\item \textbf{Matthews Correlation Coefficient (MCC)}:
\begin{equation}
    \text{MCC} = \frac{TP \cdot TN - FP \cdot FN}{\sqrt{(TP + FP)(TP + FN)(TN + FP)(TN + FN)}}
\end{equation}

\item \textbf{Geometric Mean (G-Mean)}:
\begin{equation}
    \text{G-Mean} = \sqrt{\text{Recall} \cdot \text{Specificity}}
\end{equation}

\item \textbf{Balanced Accuracy}:
\begin{equation}
    \text{Balanced Accuracy} = \frac{1}{2} \left( \text{Recall} + \text{Specificity} \right)
\end{equation}

\item \textbf{F2-Score}:
\begin{equation}
    \text{F2-Score} = \frac{5 \times \text{Precision} \times \text{Recall}}{(4 \times \text{Precision}) + \text{Recall}}
\end{equation}
\end{itemize}

\end{document}